\documentclass[12pt, draftclsnofoot, onecolumn]{IEEEtran}
\usepackage{amsfonts}
\usepackage{amssymb}
\usepackage{amsthm}
\usepackage{amsmath,amsfonts,amssymb}
\usepackage[dvips]{graphicx}
\usepackage{verbatim}
\usepackage{setspace}
\usepackage{bm}
\usepackage{extarrows}
\usepackage{algorithmic} 
\usepackage[ruled,vlined]{algorithm2e}
\usepackage{cite}

\usepackage{changepage}
\usepackage{pdfpages}
\usepackage{color}
\usepackage{bbding}

\IEEEoverridecommandlockouts
\usepackage{etoolbox}
\let\mybibitem\bibitem
\renewcommand{\bibitem}[1]{%
	\ifstrequal{#1}{nature}
	{\color{blue}\mybibitem{#1}}
	{\color{black}\mybibitem{#1}}%
}
\usepackage{subfigure}
\usepackage{graphbox}

\begin{document}
	%
	\title{\huge 3-D Positioning and Resource Allocation  for Multi-UAV Base Stations Under Blockage-Aware Channel Model}
	
	%
	%
	%
	\author{Pengfei Yi,~\IEEEmembership{Student Member,~IEEE,} 
		Lipeng Zhu,~\IEEEmembership{Member,~IEEE,}
		Zhenyu Xiao,~\IEEEmembership{Senior Member,~IEEE,}
		Rui Zhang,~\IEEEmembership{Fellow,~IEEE,}
		Zhu Han,~\IEEEmembership{Fellow,~IEEE,}
		and Xiang-Gen Xia,~\IEEEmembership{Fellow,~IEEE}
		
		\thanks{This work was supported in part by the National Key Research and Development Program under grant number 2020YFB1806800, the National Natural Science Foundation of China (NSFC) under grant numbers U22A2007, 62171010 and 61827901, and the Beijing Natural Science Foundation under grant number L212003. The corresponding author is Dr. Zhenyu Xiao with Email xiaozy@buaa.edu.cn.}
		\thanks{P. Yi and Z. Xiao are with the School of Electronic and Information Engineering, Beihang University, Beijing 100191, China. (yipengfei@buaa.edu.cn, xiaozy@buaa.edu.cn).}
		\thanks{L. Zhu is with the Department of Electrical and Computer Engineering, National University of Singapore, Singapore 117583, Singapore. (zhulp@nus.edu.sg).}
		\thanks{R. Zhang is with The Chinese University of Hong Kong, Shenzhen, and Shenzhen Research Institute of Big Data, Shenzhen, China 518172 (rzhang@cuhk.edu.cn). He is also with the Department of Electrical and Computer Engineering, National University of Singapore, Singapore 117583 (elezhang@nus.edu.sg).}
		\thanks{Z. Han is with the Department of Electrical and Computer Engineering
			in the University of Houston, Houston, TX 77004 USA, and also with the Department of Computer Science and Engineering, Kyung Hee University,
			Seoul, South Korea, 446-701. (hanzhu22@gmail.com).}
		\thanks{X.-G. Xia is with the Department of Electrical and Computer	Engineering, University of Delaware, Newark, DE 19716, USA. (xianggen@udel.edu).}
	}

	%
	%

\maketitle

\begin{abstract}
	In this paper, we propose to deploy  multiple unmanned aerial vehicle (UAV) mounted base stations to serve ground users in outdoor environments with obstacles. In particular, the geographic information is employed to capture the blockage effects for air-to-ground (A2G) links  caused by buildings, and a realistic blockage-aware A2G channel model is proposed to characterize the continuous variation of the channels at different locations. Based on the proposed channel model, we formulate the joint optimization problem of UAV three-dimensional (3-D) positioning and resource allocation, by power allocation, user association, and subcarrier allocation, to maximize the minimum achievable rate among users.  To solve this  non-convex  combinatorial programming problem,  
	we introduce a penalty term to relax it and develop a suboptimal solution via a penalty-based double-loop iterative optimization framework.  The inner loop solves the penalized problem by employing the block successive convex approximation (BSCA) technique,  where the UAV  positioning and resource allocation  are alternately optimized in each iteration.  The outer loop aims to obtain proper
	penalty multipliers to ensure the solution of the penalized problem converges to that of the original problem.
	Simulation results demonstrate the superiority of the proposed algorithm over  other benchmark schemes in terms of the minimum achievable rate.
	
\end{abstract}

\begin{IEEEkeywords}
	UAV communication, geographic information,  3-D positioning, resource allocation, blockage.
\end{IEEEkeywords}

%
\IEEEpeerreviewmaketitle

\section{Introduction}
\IEEEPARstart{I}{n} recent years, unmanned aerial vehicle (UAV)-assisted communication systems have attracted increasing attention for supporting the seamless coverage in the beyond  fifth-generation (B5G) and sixth-generation (6G) networks~\cite{mozaffari2019atutor,zeng2019access,xiao2022asurve,geraci2022whatwi}. Owing to their controllable three-dimensional (3-D) mobility and low cost, UAVs can serve as aerial base stations (BSs), relays, or access points for coverage enhancement~\cite{wu2018jointt,zhu2022multiu, liu2019optimi}, communication relaying~\cite{yi2022joint3, zhu2020millim,hu2020lowcom}, and data broadcast/collection~\cite{liu2021resour,cai2022trajec, xiao2020uavcom}. Compared to conventional terrestrial communications with typically fixed infrastructures, UAV-assisted systems offer new degrees of freedom in the spatial domain to further improve communication performance by exploiting the flexible 3-D mobility of UAVs. 

In this regard, several interesting topics arise in the study of UAV-assisted communications, such as UAV placement, trajectory design, and resource allocation.
In particular, with the increasing number of users, their wider distribution and various communication requirements make it necessary to employ multiple UAVs forming a cooperative network to improve the access capability, enlarge the coverage area, and enhance the communication reliability. In~\cite{lyu2017placem}, a successive UAV placement strategy was proposed to minimize the number of required UAVs while satisfying the communication requirement of ground users. The authors in~\cite{yin2020resour} studied the joint optimization for multi-UAV placement, user association, and resource allocation, to maximize the downlink sum rate. Multi-UAV placement and user association were also considered in~\cite{qiu2020multip}, with a particular consideration on constrained backhaul links.
Aiming to support energy-efficient Internet of Things (IoT) communications, multiple UAVs were deployed and the communication resources were jointly optimized, to minimize the transmit power of IoT devices in~\cite{liu2021resour}. Combining millimeter-wave (mmWave) communications with UAV-assisted systems, the authors in~\cite{zhu2022multiu} investigated the joint optimization of UAV placement, user clustering, and transmit/receive beamforming. In addition to the  placement optimization, there are also many works on the trajectory design for UAV communication systems, aiming at ubiquitous coverage~\cite{wu2018jointt}, secure communications~\cite{cai2018dualua}, interference coordination~\cite{shen2020multiu}, energy-efficient content coverage~\cite{zhao2021multiu}, and mobile Internet of vehicles~\cite{liu2022jointc}.


Note that the aforementioned works significantly rely on the simplified/statistical channel models for air-to-ground (A2G) communication links between UAVs and ground users. For example, the A2G channels are assumed to be dominated by  line-of-sight (LoS) paths in~\cite{lyu2017placem, wu2018jointt,yin2020resour, cai2018dualua,shen2020multiu,liu2022jointc}. While the existence of LoS links is probabilistically modeled
as a function of  the elevation angle of the A2G link in~\cite{liu2021resour,zhao2021multiu,qiu2020multip}, known as \emph{probabilistic LoS channels}~\cite{zeng2019access}. 
The simplified LoS channel models can make the positioning optimization more tractable and are suitable for average performance analysis in UAV communications. However, in practice, the terrain conditions, such as buildings and other obstacles, may cause severe blockage  to A2G links and sharply weaken the strength of the received signals, especially for dense urban areas~\cite{bai2014analys}. In such  cases,  the communication design based on a simplified LoS channel and/or statistical channel cannot guarantee the performance under site-specific   environments and may not be suitable for practical UAV-assisted communications. 

To overcome the drawback of the over-simplified channel models and capture practical propagation conditions, there are emerging research directions that exploit  two kinds of information, namely \emph{radio map} and \emph{geographic information}, for UAV-assisted communications. Constructed by a large number of real-life channel measurements, a radio map can precisely describe the average signal strength for all combinations of UAV-user locations~\cite{chen2017learni}.
Based on the radio map, the joint optimization of UAV positioning, user association, and wireless backhaul capacity allocation for a multi-UAV relay network was studied in~\cite{liu2019optimi}. In~\cite{hu2020lowcom}, the UAV trajectory and resource allocation were jointly optimized to ensure the fairness among users for a single UAV relay system. Besides, a radio map was utilized to evaluate the A2G link quality for the UAV trajectory design in a cellular-connected UAV system~\cite{zhang2021radiom} and UAV anti-jamming communications~\cite{dong2022radiom}. 
Radio map is theoretically appealing to provide precise channel quality, but it encounters difficulty in obtaining sufficient real-life channel measurements for radio map construction. Besides, a large dataset is required for reconstructing a radio map, which results in high overhead on storage and computation. In addition to radio map, geographic information is also helpful for capturing practical propagation conditions. With the available location and size information of the buildings, the existence of a LoS channel can be inferred by evaluating whether the A2G link is blocked by buildings, instead of being modeled as a random event~\cite{zeng2019access}. Based on geographic information, the authors in~\cite{zhao2020effici} proposed a geometric analysis method to detect the blockage in a multi-UAV mmWave communication system, and then developed a greedy user scheduling algorithm to decrease the probability of the blockage. In~\cite{cai2022trajec}, building blockage was considered for the trajectory design and resource allocation for a UAV-enabled data collection system. In~\cite{yi2022joint3}, the blocked regions of the ground users with respect to (w.r.t.)  buildings were modeled as polyhedrons. By restricting the UAV to be deployed  outside all the blocked regions, the LoS links can be guaranteed for a UAV relay system.

Motivated by the above works, in this paper, we study the joint optimization of UAV 3-D positioning, power allocation, user association, and subcarrier allocation to maximize the minimum achievable rate among multiple users in the downlink of multi-UAV orthogonal frequency division multiple access (OFDMA) communication systems, under the blockage-aware A2G channel model with the aid of geographic information. Different from the works based on deterministic/probabilistic LoS channel~\cite{lyu2017placem, yin2020resour,liu2021resour,qiu2020multip}, 
the geographic information utilized in this paper can predict the channel condition with a high precision. Compared to the works based on radio map~\cite{liu2019optimi, hu2020lowcom}, this work only needs the geographic information, which is easier to acquire in reality.  For example, geographic information can be derived offline from digital maps~\cite{kim2018intera}, or constructed online by using photogrammetry techniques~\cite{zhou2013comple}. 
The main contributions of this paper are summarized as follows:
\begin{enumerate}
	\item We propose to deploy multiple UAV BSs to serve multiple ground users via OFDMA. Geographic information
	is utilized to capture the realistic propagation environment. Specifically, we develop a blockage-aware A2G channel model, where the LoS and Non-LoS (NLoS) channels can be uniformly expressed as a continuous function of the normalized distance between the UAV and the ground region. Then, to ensure fairness, we formulate an optimization problem to maximize the minimum achievable rate among the users by jointly designing the 3-D positioning and resource allocation, including power allocation, user association, and subcarrier allocation.
	\item The formulated optimization problem is non-convex and involves combinatorial programming variables, and thus it is difficult to obtain the globally optimal solution. Therefore, we develop a penalty-based double-loop iterative optimization (PDLIO) algorithm to solve the problem  suboptimally. Specifically, the original problem is transformed to a penalized problem by relaxing the binary association variables into continuous ones and introducing a penalty component to  the objective function. 
	The inner-loop partitions the penalized problem into a UAV positioning sub-problem and a resource allocation sub-problem, which are alternately solved by employing the block successive convex approximation (BSCA) technique~\cite{Razaviyayn2013aunifi, yang2020inexact}. The outer-loop updates the penalty multipliers to ensure the solution of the penalized problem converges to that of the original problem. 
	\item The performance of the proposed solution for the  joint 3-D positioning and resource allocation problem in geographic information-aided multi-UAV systems is evaluated with different settings. Simulation results illustrate the convergence of the developed PDLIO algorithm and reveal performance superiority over other benchmark schemes in terms of the minimum achievable rate.
\end{enumerate}

The rest of this paper is organized as follows. In Section II, we introduce the system model, propose the blockage-aware A2G channel model based on geographic information, and formulate the joint positioning and resource allocation problem. The problem transformation and the proposed PDLIO algorithm are given in Section III. 
Section IV presents the simulation results. Finally, the paper is concluded in Section V.

\textit{Notation}: $a$, $\mathbf{a}$, $\mathbf{A}$, and $\mathcal{A}$ denote a scalar, a vector, a matrix, and a set, respectively. 
$[\mathbf{a}]_k$ denotes the $k$-th entry of vector $\mathbf{a}$.
$\|\mathbf{a}\|$ represents the Euclidean norm of vector $\mathbf{a}$. $(\cdot)^{\rm{T}}$ denotes transpose. 
$\mathbb{R}^M$ denotes the space of the $M$-dimensional real vector. 
$\mathcal{A}_1 \bigcup \mathcal{A}_2$ represents the union of sets $\mathcal{A}_1$ and $\mathcal{A}_2$.
$\mathcal{A}_2 \setminus \mathcal{A}_1$ represents the elements of $\mathcal{A}_2$ that are not included in $\mathcal{A}_1$. $|\mathcal{A}|$ denotes the cardinality of set $\mathcal{A}$. 
$\overrightarrow{AB}$ denotes the vector from point $A$ to point $B$. $\overrightarrow{AB} \cdot \overrightarrow{CD}$ and $\overrightarrow{AB} \times \overrightarrow{CD}$ denote the inner product and outer product between vector $\overrightarrow{AB}$ and vector $\overrightarrow{CD}$, respectively. For a multivariate function $f(\mathbf{x}, \mathbf{y})$, $\nabla_{\mathbf{x}} f(\mathbf{x}, \mathbf{y})$ denotes its gradient w.r.t. $\mathbf{x}$. $f(\mathbf{x}; \mathbf{x}^{l})$ denotes a surrogate function of $f(\mathbf{x})$ constructed at local point $\mathbf{x}^{{l}}$.

\section{System Model and Problem Formulation}\label{sec_systemModel}

\begin{figure}[t]
\begin{center}
	\includegraphics[width=0.6	\linewidth]{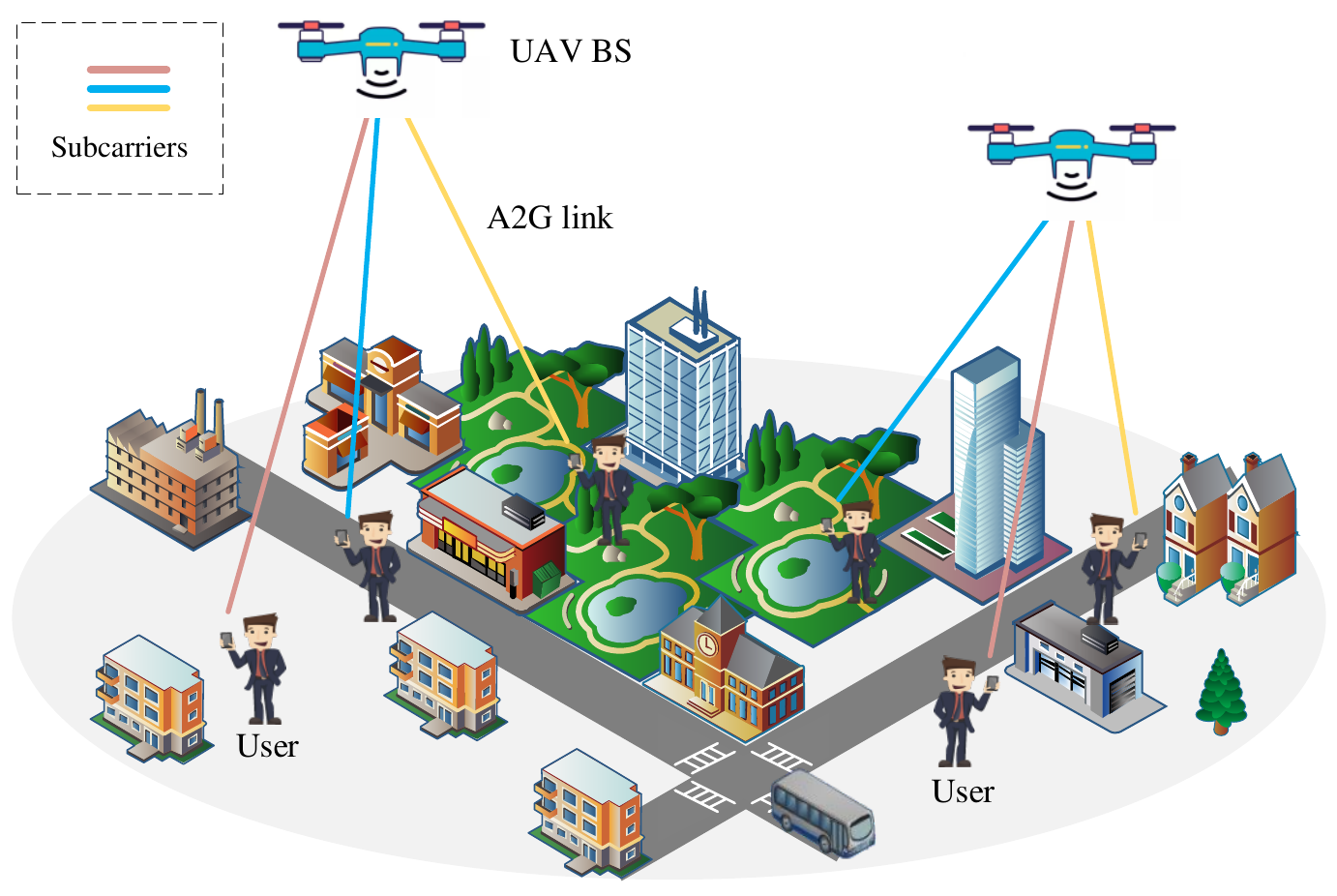}
	\caption{ Illustration of the  considered multi-UAV OFDMA downlink communication network.}
	\label{fig:multiUAVBS}
\end{center}
\end{figure}

As shown in Fig.~\ref{fig:multiUAVBS}, we consider a multi-UAV OFDMA downlink communication network in outdoor environments with obstacles, such as buildings, where $M$ UAV BSs are deployed to serve $K$ ground users via $N$ orthogonal subcarriers for each UAV. The sets of UAVs, users, and subcarriers are denoted as $\mathcal{M} \triangleq \{1,...,M\}$, $\mathcal{K} \triangleq \{1,...,K\}$, and $\mathcal{N} \triangleq \{1,...,N\}$, respectively. We assume that each UAV allocates
orthogonal subcarriers to its served ground users. As a result, the interference among these users served by the same UAV  can be eliminated. However, since all the UAVs share the same $N$ subcarriers, the users served by different UAVs via the same subcarrier may suffer from mutual interference.

Without loss of generality, we employ a 3-D Cartesian coordinate system. For user $k$, its coordinates are denoted by $\mathbf{u}_k \in \mathbb{R}^3, k \in \mathcal{K}$. The coordinates of  UAV $m$ are given by $\mathbf{x}_m \in \mathbb{R}^3, m \in \mathcal{M}$. We denote $\mathbf{X}=\left\{ \mathbf{x}_m|  m\in \mathcal{M} \right\}$. 
$Q$ buildings indexed by $\mathcal{Q}\triangleq \{1,...,Q\}$ are randomly distributed in the considered area, whose 3-D locations and sizes are assumed to be available with the aid of geographic information\footnote{A possible way to extract building information is to use an open source geographic database \emph{OpenStreetMap}, where the contour and height of a building are given by its raw data tagged with \emph{geometry} and	\emph{height}, respectively~\cite{kang2018buildi}.}. 

\subsection{Channel Model}
Note that the blockage effects caused by buildings may significantly deteriorate the link quality, which should be considered in the channel modeling.
Since the UAV positioning is designed in a relatively large timescale  compared to the channel small-scale  variation, we mainly focus on the large-scale channel characteristics. 
The channel gain between user $k$ and UAV $m$ is modeled as a function of UAV position $\mathbf{x}_m$, i.e., 
\begin{equation}\label{eq_channel}
g_k(\mathbf{x}_m)=\frac{\beta_k(\mathbf{x}_m)}{\|\mathbf{x}_m-\mathbf{u}_k\|^{\alpha_k(\mathbf{x}_m)}},
\end{equation}
where $\alpha_k(\mathbf{x}_m)$ is the path loss exponent, and $\beta_k(\mathbf{x}_m)$ is the channel gain at reference distance of 1 meter (m). Both $\alpha_k(\mathbf{x}_m)$ and $\beta_k(\mathbf{x}_m)$ are  functions of UAV position $\mathbf{x}_m$ and user position $\mathbf{u}_k$, which characterize the specific propagation conditions for A2G links (LoS and/or NLoS channels) and are defined as
\begin{equation} \label{eq_channel_parameters} 
	\begin{aligned}
		\left( \alpha_k(\mathbf{x}_m), \beta_k(\mathbf{x}_m) \right)  =
		\begin{cases}
			\left( \alpha_1, \beta_1 \right), \text{for LoS channel}, \\
			\left(  \alpha_2, \beta_2 \right), \text{for NLoS channel}.
		\end{cases}	
	\end{aligned}
\end{equation}

\begin{figure}[t]
	\begin{center}
		\includegraphics[width=0.6\linewidth]{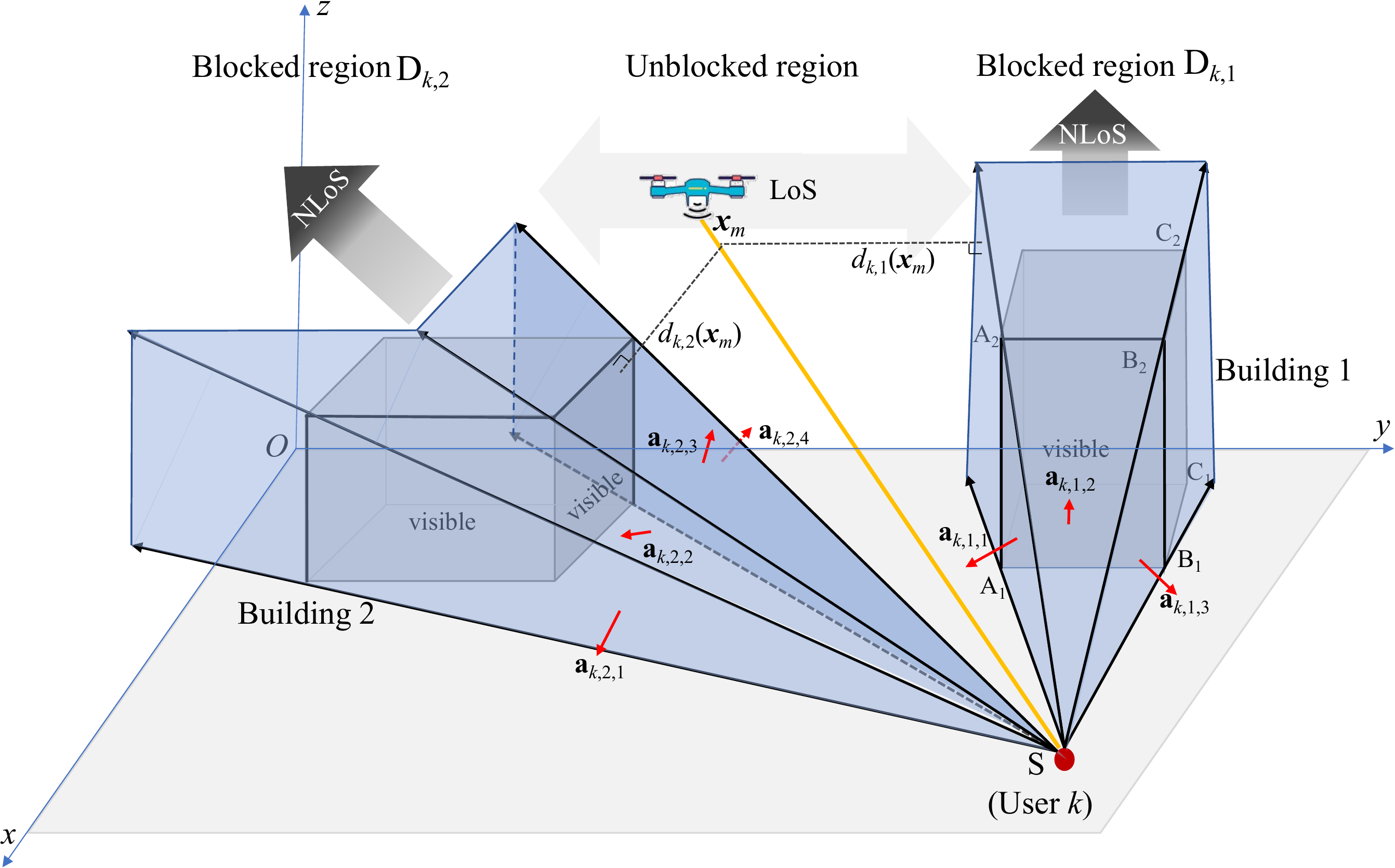}
		\caption{Illustration of the blockage caused by building.}
		\label{fig:blockage}
	\end{center}
\end{figure}

To distinguish  LoS and NLoS channels for A2G links, the key is to model the blockage effect between the UAV and user. The blocked region of user $k$ w.r.t. building $q$, denoted by $\mathcal{D}_{k,q}$, can be  modeled as a polyhedron
\begin{equation}
\mathcal{D}_{k,q}=\{ \mathbf{x} \in \mathbb{R}^3 | \mathbf{a}_{k,q,i}^{\rm{T}} \mathbf{x} -b_{k,q,i} \le 0,  i\in \mathcal{I}_{k,q} \},
\end{equation}
where $\mathcal{I}_{k,q}$ is the set of indices of hyperplanes (boundaries) for blocked region $\mathcal{D}_{k,q}$. 
$\mathbf{a}_{k,q,i} \in \mathbb{R}^3$ and $b_{k,q,i}\in \mathbb{R}$ are the outward normal vector and offset of the $i$-th hyperplane, respectively, which are determined by the position of the user and the visible flank surfaces of the building. 
The detailed procedures to identify visible flank surfaces of the building and determine the boundaries of the blocked region can be referred to~\cite[Algorithm~1]{yi2022joint3}. 
Here, we provide a toy example of an area with two buildings for illustration. As shown in Fig.~\ref{fig:blockage}, the flank surface $A_1B_1B_2A_2$ of the first building  is visible for user $k$, and thus three hyperplanes, $SA_1A_2$, $SA_2B_2$, and $SB_2B_1$, form the boundaries of blocked region $\mathcal{D}_{k,1}$. The outward normal vector is obtained as $\mathbf{a}_{k,1,1}=\frac{\overrightarrow{SA_2} \times \overrightarrow{SA_1}}{\|\overrightarrow{SA_2} \times \overrightarrow{SA_1}\|}$, and the corresponding offset w.r.t. the origin $O$ is given by $b_{k,1,1}=\mathbf{a}_{k,1,1}^{\rm{T}} \cdot \overrightarrow{OS}$, where $O$ denotes the origin  of the coordinate system. The blocked region $\mathcal{D}_{k,2}$ can be obtained in a similar manner, which is surrounded by four hyperplanes.

With the union of the obtained blocked regions,   $\mathop{\bigcup}_{q\in \mathcal{Q}} \mathcal{D}_{k,q}$, of user $k$, the existence of LoS path  between user $k$ and UAV $m$ can be inferred directly by evaluating whether the UAV is located in the blocked regions, $\mathop{\bigcup}_{q\in \mathcal{Q}} \mathcal{D}_{k,q}$. If UAV $m$ is deployed in any of the blocked regions of user $k$, i.e., 
$\mathbf{x}_m \in \mathop{\bigcup}_{q\in \mathcal{Q}} \mathcal{D}_{k,q}$, 
the LoS path between them is blocked. Otherwise, for $\mathbf{x}_m \notin  \mathop{\bigcup}_{q\in \mathcal{Q}} \mathcal{D}_{k,q}$, the LoS channel is guaranteed without any building blockage. Define
\begin{equation}
d_{k,q}(\mathbf{x}_m)\triangleq \max\limits_{i\in \mathcal{I}_{k,q}} \left\{  \mathbf{a}_{k,q,i}^{\rm{T}} \mathbf{x}_m -b_{k,q,i}  \right\},
\end{equation}
and then $\mathbf{x}_m \in \mathcal{D}_{k,q}$ is equivalent to $d_{k,q}(\mathbf{x}_m)\leq 0$. 
Therefore, $\mathbf{x}_m \in \mathop{\bigcup}_{q\in \mathcal{Q}} \mathcal{D}_{k,q}$ (NLoS condition) is equivalent to 
\begin{equation}
\min\limits_{q\in \mathcal{Q}} \left\{ d_{k,q}(\mathbf{x}_m) \right\} \leq 0, \notag
\end{equation}
while $\mathbf{x}_m \notin \mathop{\bigcup}_{q\in \mathcal{Q}} \mathcal{D}_{k,q}$ (LoS condition) is equivalent to
\begin{equation}
\min\limits_{q\in \mathcal{Q}} \left\{ d_{k,q}(\mathbf{x}_m) \right\} > 0. \notag
\end{equation}
As can be observed, the LoS/NLoS condition can be described as a \emph{step function} w.r.t. $\min_{q\in \mathcal{Q}}$ $\left\{ d_{k,q}(\mathbf{x}_m) \right\}$. For the example shown in Fig.~\ref{fig:blockage}, we have $\min_{q\in \{1,2\}} \left\{ d_{k,q}(\mathbf{x}_m) \right\} > 0$, because of $d_{k,1}(\mathbf{x}_m)= \max\limits_{i\in \mathcal{I}_{k,1}} \left\{  \mathbf{a}_{k,1,i}^{\rm{T}} \mathbf{x}_m -b_{k,1,i} \right\}=\mathbf{a}_{k,1,1}^{\rm{T}} \mathbf{x}_m -b_{k,1,1}>0$ and $d_{k,2}(\mathbf{x}_m)= \max\limits_{i\in \mathcal{I}_{k,2}} \left\{  \mathbf{a}_{k,2,i}^{\rm{T}} \mathbf{x}_m -b_{k,2,i} \right\}$ $=\mathbf{a}_{k,2,3}^{\rm{T}} \mathbf{x}_m -b_{k,2,3}>0$. Therefore, the A2G link between UAV $m$ and user $k$ satisfies the LoS condition.

In order to facilitate the subsequent optimization procedure, the channel model provided in~\eqref{eq_channel} and~\eqref{eq_channel_parameters} needs to be formulated as a continuous and differentiable function  w.r.t. the position of the UAV. Therefore, we employ the \emph{sigmoid function} to approximate the step function, which is given by
\begin{align}\label{smooth function}
s(\mathbf{x}_m,\mathbf{u}_k)=\frac{1}{1+\exp \left(-\eta \frac{\min_{q\in \mathcal{Q}} \left\{ d_{k,q}(\mathbf{x}_m)\right\}}{\|\mathbf{x}_m-\mathbf{u}_k\|}  \right)},
\end{align}
where the smooth parameter $\eta$ can be used to control the scale of the approximation.  From a physical point of view, $\eta$ characterizes the change rate between different degrees of channel blockage, including LoS, obstructed LoS, and NLoS~\cite{chen20213durba}. The distance of the A2G link, $\|\mathbf{x}_m-\mathbf{u}_k\|$, is used for normalization, which guarantees an identical channel blockage condition along the same direction relative to the user position. Thus, the channel parameters $\alpha_k(\mathbf{x}_m)$ and $\beta_k(\mathbf{x}_m)$  in \eqref{eq_channel_parameters} can be approximated as   
\begin{align}\label{eq_channel_parameter}
\left\{\begin{matrix}
	\alpha_k(\mathbf{x}_m)=(\alpha_1-\alpha_2) \cdot s(\mathbf{x}_m,\mathbf{u}_k)  +\alpha_2, 	\\
	\beta_k(\mathbf{x}_m)=(\beta_1-\beta_2)\cdot s(\mathbf{x}_m,\mathbf{u}_k)  +\beta_2. 
\end{matrix}\right.
\end{align}

By substituting~\eqref{eq_channel_parameter} into \eqref{eq_channel}, we then obtain a blockage-aware A2G channel model, which involves position-dependent channel parameters to distinguish in different propagation environments.
In particular, for $\eta \to \infty$, we have 
\begin{equation}\label{eq_channel_LoSNLoS} %
\lim\limits_{^{\eta \to \infty}} g_k(\mathbf{x}_m)=
\begin{cases}
	\frac{\beta_1}{\|\mathbf{x}_m-\mathbf{u}_k\|^{\alpha_1}},\min\limits_{q\in \mathcal{Q}} \left\{ d_{k,q}(\mathbf{x}_m) \right\} \geq \Delta, \\
	\frac{\beta_2}{\|\mathbf{x}_m-\mathbf{u}_k\|^{\alpha_2}},\min\limits_{q\in \mathcal{Q}} \left\{ d_{k,q}(\mathbf{x}_m)\right\} \leq -\Delta,
\end{cases}
\end{equation}
where $\Delta>0$ is an arbitrarily small positive constant. The model in \eqref{eq_channel_LoSNLoS} is consistent with the traditional segmented channel models which classify
the propagation into LoS and NLoS conditions~\cite{chen2017learni, zeng2019access}.



\subsection{Problem Formulation}
Denote $\mathbf{X}=\left\{ \mathbf{x}_{m},  m\in \mathcal{M}  \right\}$ as the collection of the positions of UAVs, and $\mathbf{P}=\left\{ p_{m,n},  m\in \right.$ $\left. \mathcal{M}, n\in \mathcal{N}  \right\}$ as the collection of the transmit powers of UAVs, in which $p_{m,n}$ is the transmit power allocated on subcarrier $n$ at UAV $m$. Assuming that user $k$ is served by UAV $m$ over subcarrier $n$, then the received signal-to-interference-plus-noise-ratio (SINR) of user $k$ on this subcarrier can be expressed as
\begin{equation}\label{eq_SINR}
\mathrm{SINR}_{k,m,n} \left(\mathbf{X}, \mathbf{P} \right)=\frac{p_{m,n}g_k(\mathbf{x}_m)}{\sum\limits_{j \in \mathcal{M}\setminus \{m\}} p_{j,n}g_k(\mathbf{x}_j)+\sigma^2},
\end{equation}
where $\sigma^2$ is the power of the additive  white Gaussian noise (AWGN) at user $k$.

Denote $\mathbf{C}=\left\{ c_{k,m,n} ,  k\in \mathcal{K},  m\in \mathcal{M},  n\in \mathcal{N}  \right\}$ as the collection of user-UAV-subcarrier association indicators. If user $k$ is served by UAV $m$ over subcarrier $n$, we have $c_{k,m,n}=1$; otherwise, $c_{k,m,n}=0$. 
Then, the achievable rate (in bits/sec/Hz) of user $k$ served by UAV $m$ over subcarrier $n$ is expressed as a function of $\mathbf{X}$, $\mathbf{P}$, and $\mathbf{C}$, i.e.,
\begin{equation}\label{eq_rate_kmn}
\begin{aligned}
	R_{k,m,n}\left(\mathbf{X}, \mathbf{P}, \mathbf{C} \right)&=c_{k,m,n} \log  \left(1+\mathrm{SINR}_{k,m,n} \left(\mathbf{X}, \mathbf{P} \right) \right) \\
	&=c_{k,m,n} \log \left(1+\frac{p_{m,n}g_k(\mathbf{x}_m)}{\sum\limits_{j \in \mathcal{M}\setminus \{m\}} p_{j,n}g_k(\mathbf{x}_j)+\sigma^2}\right),
\end{aligned}
\end{equation}
with the base of $\log$ equal to 2.
Then, the achievable rate of user $k$ is given by
\begin{equation}\label{eq_rate}
\begin{aligned}
	R_{k}(\mathbf{X},\mathbf{P}, \mathbf{C})&=\sum_{m\in \mathcal{M}} \sum_{n\in \mathcal{N}} R_{k,m,n}\left(\mathbf{X}, \mathbf{P}, \mathbf{C} \right). \\
\end{aligned}
\end{equation}

To maximize the minimum	achievable rate among all the ground users, we formulate the following problem for optimizing the UAV positioning $\mathbf{X}$, power allocation $\mathbf{P}$, and user association along with subcarrier allocation $\mathbf{C}$:
\begin{align}
\max\limits_{\mathbf{X},\mathbf{P}, \mathbf{C}}~~ & \min\limits_{ k\in \mathcal{K}} R_{k}(\mathbf{X},\mathbf{P}, \mathbf{C})     \label{eq_problem_MultiUAV}\\
\mbox{s.t.}~~ 	&c_{k,m,n}\in \{0, 1\},  k\in \mathcal{K},  m\in \mathcal{M},  n\in \mathcal{N}, \tag{\ref{eq_problem_MultiUAV}a}\label{c:ckmn} \\
&\sum_{k\in \mathcal{K}} c_{k,m,n} \leq 1,  m\in \mathcal{M},  n\in \mathcal{N},  \tag{\ref{eq_problem_MultiUAV}b}\label{c:orthogonal} \\
&\sum_{m\in \mathcal{M}} \sum_{n\in \mathcal{N}} c_{k,m,n} = 1,   k\in \mathcal{K}, \tag{\ref{eq_problem_MultiUAV}c}\label{c:onetoone} \\
&p_{m,n} \geq 0,  m\in \mathcal{M}, n\in \mathcal{N},  \tag{\ref{eq_problem_MultiUAV}d}\label{c:power0}  \\
&\sum_{n\in \mathcal{N}} p_{m,n} \leq P_{\mathrm{max}},  m\in \mathcal{M}, \tag{\ref{eq_problem_MultiUAV}e}\label{c:power_max} \\
& \mathbf{x}_m \in \mathcal{D},  m\in \mathcal{M}, \tag{\ref{eq_problem_MultiUAV}f}\label{c:region}
\\
& \|\mathbf{x}_m -\mathbf{x}_j\|^2  \geq d_\mathrm{min}^2,  m,j\in \mathcal{M}, m\neq j, \tag{\ref{eq_problem_MultiUAV}g}\label{c:secure_dis}
\end{align}
where $\min_{ k\in \mathcal{K}} R_{k}(\mathbf{X},\mathbf{P}, \mathbf{C})$
denotes the minimum achievable rate
among all the users. Constraint \eqref{c:ckmn} indicates that $c_{k,m,n}$ is binary.
Constraint \eqref{c:orthogonal} ensures that on each subcarrier, each UAV can serve one user at most. Constraint \eqref{c:onetoone} guarantees that each user can be connected to  one UAV over one subcarrier.
Constraints \eqref{c:power0} and \eqref{c:power_max} indicate that the transmit power of each UAV 
on each subcarrier is nonnegative and the total transmit power does not exceed a maximum value $P_\mathrm{max}$.
Constraint \eqref{c:region} confines the region that each UAV can be positioned, where 
$\mathcal{D}=\{ \mathbf{x}\in \mathbb{R}^3 | [\mathbf{x}]_1 \in [0,x_\mathrm{D}], [\mathbf{x}]_2\in [0,y_\mathrm{D}], [\mathbf{x}]_3 \geq h_\mathrm{min} \}$ denotes the whole considered region.
Finally, in practice, the positions of UAVs are subject to the collision avoidance constraint \eqref{c:secure_dis}, where $d_\mathrm{min}$ denotes the minimum inter-UAV distance to ensure collision avoidance.
As can be seen, problem~\eqref{eq_problem_MultiUAV} involves combinatorial programming variables, 
and the variables are highly coupled. It is challenging to obtain the globally optimal solution for this problem. To address this issue,  we propose a suboptimal solution for problem~\eqref{eq_problem_MultiUAV} in the following section.

\section{Proposed Solution}\label{sec_solution_relaxed}
In this section, we propose to employ the penalty method and BSCA technique to solve  the  optimization problem~\eqref{eq_problem_MultiUAV}. In Section~\ref{sub_Sec_Relaxed}, we first transform problem~\eqref{eq_problem_MultiUAV} to a penalized problem, where the binary variables $\mathbf{C}$ are relaxed to continuous ones and a penalty term is introduced to the objective function. Then, a double-loop optimization framework is developed. The inner-loop solves the penalized problem for given penalty multipliers  by alternately optimizing UAV positioning $\mathbf{X}$ and resource allocation $\{\mathbf{P}, \mathbf{C}\}$, which are introduced in Sections~\ref{sub_Sec_Positioning} and \ref{sub_Sec_RA}, respectively. The outer-loop updates the penalty multipliers to decrease the violation of the relaxed constraints, as detailed  in Section~\ref{sub_Sec_Multiplier}. We finally present the overall PDLIO algorithm in Section~\ref{sub_Sec_Overall}.

\subsection{ Problem Transformation}\label{sub_Sec_Relaxed}
To make the binary constraint in~\eqref{c:ckmn} more tractable, we replace it equivalently by
\begin{align}
&0 \leq c_{k,m,n} \leq 1,   k\in \mathcal{K},  m\in \mathcal{M},  n\in \mathcal{N}, \label{c:c_kmn_continuous} \\
& c_{k,m,n}(1-c_{k,m,n})\leq 0,   k\in \mathcal{K},  m\in \mathcal{M},  n\in \mathcal{N}. \label{c:c_kmn_continuous2}
\end{align}	
In this way, $c_{k,m,n}$ becomes a continuous optimization variable between 0 and 1. By  penalizing constraint \eqref{c:c_kmn_continuous2} into the objective function with multipliers $\mathbf{\Lambda}=\{\lambda_{k,m,n},  k\in \mathcal{K},  m\in \mathcal{M},  n\in \mathcal{N}\}$, we obtain a  penalized problem
\begin{align}
\max\limits_{\mathbf{X},\mathbf{P}, \mathbf{C}}~~ & \min\limits_{ k\in \mathcal{K}} R_{k}(\mathbf{X},\mathbf{P}, \mathbf{C}) + \rho(\mathbf{\Lambda}, \mathbf{C})  \label{eq_problem_relaxed}\\
\mbox{s.t.}~~ 	&\lambda_{k,m,n} \geq 0,  k\in \mathcal{K},  m\in \mathcal{M}, n\in \mathcal{N}, \tag{\ref{eq_problem_relaxed}a}\label{c:lambda}  \\
&\eqref{c:orthogonal},\eqref{c:onetoone}, \eqref{c:power0}, \eqref{c:power_max},\eqref{c:region}, \eqref{c:secure_dis},\eqref{c:c_kmn_continuous}, \notag
\end{align}
where $\rho(\mathbf{\Lambda}, \mathbf{C}) \triangleq -\sum\limits_{k \in \mathcal{K}}\sum\limits_{m \in \mathcal{M}}\sum\limits_{n \in \mathcal{N}} \lambda_{k,m,n}  c_{k,m,n}(1-c_{k,m,n})$. 

Note that multiplier $\lambda_{k,m,n}$ avoids the violation of constraint~\eqref{c:c_kmn_continuous2}. In particular, for $\lambda_{k,m,n}\to \infty,  k\in \mathcal{K},  m\in \mathcal{M},  n\in \mathcal{N}$,  problem~\eqref{eq_problem_relaxed} is equivalent to the original problem~\eqref{eq_problem_MultiUAV}~\cite{bertsekas1997nonlinear}, and thus they have the same optimal solution. However,  problem~\eqref{eq_problem_relaxed} is still non-convex and may be suboptimally solved during the iterations. In such case, 
it is not wise  to initialize $\lambda_{k,m,n}$  to be too  large, since the objective will be dominated by the penalty term $\rho(\mathbf{\Lambda}, \mathbf{C})$ and the minimum achievable rate term $\min_{ k\in \mathcal{K}} R_{k}(\mathbf{X},\mathbf{P}, \mathbf{C})$ will be diminished. Therefore, we initialize $\lambda_{k,m,n}$ as a small value to provide enough degrees of freedom for UAV positioning and resource allocation to obtain a good solution. Then, by gradually increasing the value of $\lambda_{k,m,n}$, the violation of constraint~\eqref{c:c_kmn_continuous2} can be gradually decreased until it is strictly satisfied, which ensures that the solution for penalized problem~\eqref{eq_problem_relaxed} converges to a feasible solution  for the original problem~\eqref{eq_problem_MultiUAV}. 

In the next two subsections, for the $(L+1)$-th outer loop with given multipliers $\Lambda^L$, we introduce a method to solve the penalized problem~\eqref{eq_problem_relaxed} through inner-loop iterations.
Define
\begin{equation}\label{eq_rate_kmn_sub}\notag
\begin{aligned}
	\hat{R}_{k,m,n}\left(\mathbf{X}, \mathbf{P}, \mathbf{C} \right)&=c_{k,m,n} \log \left(1+\sum_{j \in \mathcal{M}}  \frac{p_{j,n}}{\sigma^2} g_k(\mathbf{x}_j) \right),\\
	\bar{R}_{k,m,n}\left(\mathbf{X}, \mathbf{P}, \mathbf{C} \right)&=c_{k,m,n} \log \left(1+\sum_{j \in \mathcal{M}\setminus \{m\}} \frac{p_{j,n}}{\sigma^2} g_k(\mathbf{x}_j) \right),
\end{aligned}
\end{equation}
and then \eqref{eq_rate_kmn} is equivalent to
\begin{equation}\label{eq_rate_kmn_copy}\notag
\begin{aligned}
	R_{k,m,n}\left(\mathbf{X}, \mathbf{P}, \mathbf{C} \right)= \hat{R}_{k,m,n}\left(\mathbf{X}, \mathbf{P}, \mathbf{C} \right) - \bar{R}_{k,m,n}\left(\mathbf{X}, \mathbf{P}, \mathbf{C} \right).
\end{aligned}
\end{equation}	
For notational simplicity, we define
$$Z(\mathbf{X}, \mathbf{P},\mathbf{C})= \min\limits_{k\in \mathcal{K}} R_{k}\left(\mathbf{X}, \mathbf{P},\mathbf{C} \right)+ \rho(\mathbf{\Lambda}^L, \mathbf{C}), $$ which is the objective of problem~\eqref{eq_problem_relaxed} with given $\Lambda^{L}$.

\subsection{UAV Positioning}\label{sub_Sec_Positioning}
For the $(l+1)$-th iteration  of the inner-loop, given the power allocation $\mathbf{P}^{{l}}=\{p_{m,n}^{{l}}\}$ and association  $\mathbf{C}^{{l}}=\{c_{k,m,n}^{{l}}\}$, problem \eqref{eq_problem_relaxed} is transformed to the following UAV positioning problem:
\begin{align}
\max\limits_{\mathbf{X}}~~ & Z(\mathbf{X}, \mathbf{P}^{{l}},\mathbf{C}^{{l}})    \label{eq_problem_positioning}\\
\mbox{s.t.}~~ 	&\eqref{c:region}, \eqref{c:secure_dis}. \notag
\end{align}

Problem~\eqref{eq_problem_positioning} is a non-convex problem because of the non-concave objective function  and non-convex  constraint \eqref{c:secure_dis}. To tackle this problem, we solve a local approximation of problem~\eqref{eq_problem_positioning} based on the BSCA technique~\cite{Razaviyayn2013aunifi, yang2020inexact}.
This method mainly involves three steps as follows:
\begin{itemize}
\item Construct a concave approximation of  $Z(\mathbf{X}, \mathbf{P}^{{l}},\mathbf{C}^{{l}})$ at local point $\mathbf{X}^{{l}}$ in the $l$-th iteration of $\mathbf{P}^{{l}}$ and $\mathbf{C}^{{l}}$;
\item Solve the local approximation problem to obtain an ascent direction of the original objective function;
\item Select a proper stepsize to update positioning variables $\mathbf{X}$ for yielding an increase of the objective value.
\end{itemize}


\subsubsection{Construction of Approximation Functions}

We aim to design a surrogate function $Z^\mathrm{Pos}\left(\mathbf{X};\right.$
$\left. \mathbf{X}^{{l}}, \mathbf{P}^{{l}},\mathbf{C}^{{l}} \right)$, which is a concave function approximation of the original objective function $Z(\mathbf{X}, \mathbf{P}^{{l}},$
$\mathbf{C}^{{l}})$ at local point $\mathbf{X}^{{l}}=\{\mathbf{x}_{m,n}^{{l}}\}$. The surrogate function $Z^\mathrm{Pos}\left(\mathbf{X}; \mathbf{X}^{{l}}, \mathbf{P}^{{l}},\mathbf{C}^{{l}} \right)$ is chosen to be a strongly concave function such that problem~\eqref{eq_problem_positioning} can be relaxed to a convex problem. Moreover, it should satisfy the following conditions to ensure the local equivalence property~\cite{Razaviyayn2013aunifi}:
\begin{align}
Z^\mathrm{Pos}\left(\mathbf{X}^{{l}}; \mathbf{X}^{{l}}, \mathbf{P}^{{l}},\mathbf{C}^{{l}} \right)&=Z\left(\mathbf{X}^{{l}}, \mathbf{P}^{{l}}, \mathbf{C}^{{l}} \right) , \label{c:value} \\
\nabla_{\mathbf{X}} Z^\mathrm{Pos}\left(\mathbf{X}^{{l}}; \mathbf{X}^{{l}}, \mathbf{P}^{{l}},\mathbf{C}^{{l}} \right) &= \nabla_{\mathbf{X}} Z\left(\mathbf{X}^{{l}}, \mathbf{P}^{{l}}, \mathbf{C}^{{l}} \right). \label{c:gradient}
\end{align}

To this end,
we first design a concave function  approximation of the channel gain $g_k(\mathbf{x}_m)$. 
Note that in practice, it is unlikely that the large-scale  propagation environment rapidly changes over a small area. Therefore, the propagation parameters $\alpha_k(\mathbf{x}_m)$ and $\beta_k(\mathbf{x}_m)$ are approximately the same over a local region around $\mathbf{x}_m^{{l}}$. Then, an approximation on  the channel gain $g_k(\mathbf{x}_m)$ in \eqref{eq_channel} can be given by 
\begin{align}\label{g_k}
g_k(\mathbf{x}_m)\approx \underbrace{\tilde{g}_k(\mathbf{x}_m;\mathbf{x}_m^{{l}} )}_\text{Predicted channel gain}+  \underbrace{\delta_k(\mathbf{x}_m^{{l}}) \left( \mathbf{x}_m-\mathbf{x}_m^{{l}} \right)}_\text{Correction term},
\end{align}
where 
$\tilde{g}_k(\mathbf{x}_m;\mathbf{x}_m^{{l}}) \triangleq \beta_k(\mathbf{x}_m^{{l}})\|\mathbf{x}_m-\mathbf{u}_k\|^{-\alpha_k(\mathbf{x}_m^{{l}})}
$ represents the predicted channel gain between UAV $m$ located at $\mathbf{x}_{m}$ and user $k$, based on the current channel state with fixed channel parameters $\alpha_k(\mathbf{x}_m^{{l}})$ and  $\beta_k(\mathbf{x}_m^{{l}})$. 
$\delta_k(\mathbf{x}_m^{{l}}) \left( \mathbf{x}_m-\mathbf{x}_m^{{l}} \right)$ acts as a correction term to rectify the prediction errors, as well as  ensures the gradient consistency, with
$\delta_k(\mathbf{x}_m^{{l}})= \nabla_{\mathbf{x}_m}^\mathrm{T} g_k\left(\mathbf{x}_m^{{l}} \right) - \nabla_{\mathbf{x}_m}^\mathrm{T} \tilde{g}_k\left(\mathbf{x}_m^{{l}};\mathbf{x}_m^{{l}}\right)$.  
Besides, $\nabla_{\mathbf{x}_m} g_k\left(\mathbf{x}_m^{{l}} \right)$ and $\nabla_{\mathbf{x}_m} \tilde{g}_k\left(\mathbf{x}_m^{{l}};\mathbf{x}_m^{{l}}\right)$ can be derived as
\begin{equation}
\begin{aligned}
	\nabla_{\mathbf{x}_m} g_k\left(\mathbf{x}_m^{{l}} \right)=&- 
	g_k\left(\mathbf{x}_m^{{l}} \right) \cdot        \nabla_{\mathbf{x}_m}\alpha_k\left(\mathbf{x}_m^{{l}} \right) \cdot \log  (\|\mathbf{x}_m^{{l}}-\mathbf{u}_k\|) 
	\\&-g_k\left(\mathbf{x}_m^{{l}} \right) \cdot\frac{\alpha_k\left(\mathbf{x}_m^{{l}} \right) \cdot (\mathbf{x}_m^{{l}}-\mathbf{u}_k) }{\|\mathbf{x}_m^{{l}}-\mathbf{u}_k\|^2}   +  \frac{\nabla_{\mathbf{x}_m}\beta_k\left(\mathbf{x}_m^{{l}} \right) }{\|\mathbf{x}_m^{{l}}-\mathbf{u}_k\|^{\alpha_k\left(\mathbf{x}_m^{{l}} \right)} }
	, \notag
\end{aligned}
\end{equation}
\begin{align}
\nabla_{\mathbf{x}_m} \tilde{g}_k\left(\mathbf{x}_m^{{l}};\mathbf{x}_m^{{l}} \right)=   -\tilde{g}_k\left(\mathbf{x}_m^{{l}};\mathbf{x}_m^{{l}} \right) \cdot \frac{\alpha_k(\mathbf{x}_m^{{l}} ) \cdot (\mathbf{x}_m^{{l}}-\mathbf{u}_k) }{\|\mathbf{x}_m^{{l}}-\mathbf{u}_k\|^2}. \notag
\end{align}
The detailed procedures to derive $\nabla_{\mathbf{x}_m} g_k\left(\mathbf{x}_m \right)$,  $\nabla_{\mathbf{x}_m}\alpha_k\left(\mathbf{x}_m \right)$, and $\nabla_{\mathbf{x}_m}\beta_k\left(\mathbf{x}_m \right)$ are provided in Appendix~\ref{Appendix_gradient}.


Note that $\tilde{g}_k(\mathbf{x}_m;\mathbf{x}_m^{{l}})$ is still not concave w.r.t.  $\mathbf{x}_m$. By taking the first-order Taylor expansion of $\tilde{g}_k(\mathbf{x}_m;\mathbf{x}_m^{{l}})$ w.r.t.  $\|\mathbf{x}_m-\mathbf{u}_k\|^2$, we can obtain a concave function  approximation of $\tilde{g}_k(\mathbf{x}_m;\mathbf{x}_m^{{l}})$ as
\begin{equation}\label{tilde_g_k}
\begin{aligned}
	\tilde{g}_k(\mathbf{x}_m;\mathbf{x}_m^{{l}}) &\approx 
	A_{k,m}^{{l}}
	\left( \|\mathbf{x}_m^{{l}}-\mathbf{u}_k\|^2-\|\mathbf{x}_m-\mathbf{u}_k\|^2 \right)+\tilde{g}_k(\mathbf{x}_m^{{l}};\mathbf{x}_m^{{l}}),
\end{aligned}
\end{equation}
with $A_{k,m}^{{l}}=\alpha_k(\mathbf{x}_m^{{l}}) \beta_k(\mathbf{x}_m^{{l}}) / \left(  2 \|\mathbf{x}_m^{{l}}-\mathbf{u}_k\|^{2+\alpha_k(\mathbf{x}_m^{{l}})}   \right)
$.


Based on the above approximation of the channel gain,  a concave function approximation of $\hat{R}_{k,m,n}\left(\mathbf{X}, \mathbf{P}^{{l}}, \mathbf{C}^{{l}} \right)$ around $\mathbf{X}^{{l}}$ can be obtained as
\begin{equation}\label{eq_R_kmn_appro1}
\begin{aligned}
	&\hat{R}_{k,m,n}\left(\mathbf{X}, \mathbf{P}^{{l}}, \mathbf{C}^{{l}} \right) \\&\overset{(a)}{\approx}   B_{k,m,n}^{{l}} \sum_{j \in \mathcal{M}}  \frac{p_{j,n}^{{l}}}{\sigma^2} \left( g_k(\mathbf{x}_j)-g_k(\mathbf{x}_j^{{l}})\right)+\hat{R}_{k,m,n}\left(\mathbf{X}^{{l}}, \mathbf{P}^{{l}},\mathbf{C}^{{l}} \right) \\& 
	\overset{(b)}{\approx} B_{k,m,n}^{{l}} \sum_{j \in \mathcal{M}}  \frac{p_{j,n}^{{l}}}{\sigma^2} \left( A_{k,j}^{{l}}
	\left( \|\mathbf{x}_j^{{l}}-\mathbf{u}_k\|^2-\|\mathbf{x}_j-\mathbf{u}_k\|^2 \right) + \delta_k(\mathbf{x}_j^{{l}}) \left( \mathbf{x}_j-\mathbf{x}_j^{{l}}\right)
	\right) + \hat{R}_{k,m,n}\left(\mathbf{X}^{{l}}, \mathbf{P}^{{l}},\mathbf{C}^{{l}} \right)	\\	 &\triangleq \hat{R}_{k,m,n}^{\mathrm{appr}}\left(\mathbf{X}; \mathbf{X}^{{l}}, \mathbf{P}^{{l}}, \mathbf{C}^{{l}} \right)
\end{aligned}
\end{equation}
with $B_{k,m,n}^{{l}}=c_{k,m,n}^{{l}}/\left(1+\sum_{j \in \mathcal{M}}  \frac{p_{j,n}^{{l}}}{\sigma^2} g_k(\mathbf{x}_j^{{l}})\right)$. The approximation in step $(a)$ is obtained according to the first-order Taylor expansion of $\log(1+y)$, and the approximation in step $(b)$ is obtained by replacing $g_k(\mathbf{x}_j)$ with its approximation in \eqref{g_k}, in which $\tilde{g}_k(\mathbf{x}_j;\mathbf{x}_j^{{l}})$ is replaced by its concave function  approximation in \eqref{tilde_g_k}.

On the other hand, a linear function  approximation of  $\bar{R}_{k,m,n}\left(\mathbf{X}, \mathbf{P}^{{l}}, \mathbf{C}^{{l}} \right)$ can be obtained as 
\begin{equation}\label{eq_R_kmn_appro2}
\begin{aligned}
	\bar{R}_{k,m,n}\left(\mathbf{X}, \mathbf{P}^{{l}}, \mathbf{C}^{{l}} \right) &\overset{(c)}{\approx}   B_{k,-m,n}^{{l}} \sum_{j \in \mathcal{M}\setminus \{m\} }  \frac{p_{j,n}^{{l}}}{\sigma^2} \left( g_k(\mathbf{x}_j)-g_k(\mathbf{x}_j^{{l}})\right)+\bar{R}_{k,m,n}\left(\mathbf{X}^{{l}}, \mathbf{P}^{{l}}, \mathbf{C}^{{l}} \right) \\& 
	\overset{(d)}{\approx} B_{k,-m,n}^{{l}} \sum_{j \in \mathcal{M}\setminus \{m\}}  \frac{p_{j,n}^{{l}}}{\sigma^2} \cdot \nabla_{\mathbf{x}_j}^\mathrm{T} g_k\left(\mathbf{x}_j^{{l}} \right) \cdot \left( \mathbf{x}_j-\mathbf{x}_j^{{l}}\right) + \bar{R}_{k,m,n}\left(\mathbf{X}^{{l}}, \mathbf{P}^{{l}}, \mathbf{C}^{{l}} \right) \\
	&\triangleq \bar{R}_{k,m,n}^{\mathrm{appr}}\left(\mathbf{X}; \mathbf{X}^{{l}}, \mathbf{P}^{{l}}, \mathbf{C}^{{l}} \right)
\end{aligned}
\end{equation}
with $B_{k,-m,n}^{{l}}=c_{k,m,n}^{{l}}/\left(1+\sum_{j \in \mathcal{M}\setminus \{m\}}  \frac{p_{j,n}^{{l}}}{\sigma^2} g_k(\mathbf{x}_j^{{l}})\right)$. The approximation in step $(c)$ is obtained according to the first-order Taylor expansion of $\log(1+y)$, and the approximation in step $(d)$ is obtained by replacing $g_k(\mathbf{x}_j)$ with its first-order Taylor expansion. 

Therefore, we can obtain a concave function approximation of $R_{k}\left(\mathbf{X}, \mathbf{P}^{{l}},\mathbf{C}^{{l}} \right)$ as 
\begin{equation}\label{eq_R_kmn_appro_final}\notag
\begin{aligned}
	R_{k}^{\mathrm{Pos}}\left(\mathbf{X}; \mathbf{X}^{{l}}, \mathbf{P}^{{l}},\mathbf{C}^{{l}} \right)&= \sum_{m\in \mathcal{M}} \sum_{n\in \mathcal{N}}  \hat{R}_{k,m,n}^{\mathrm{appr}}\left(\mathbf{X}; \mathbf{X}^{{l}}, \mathbf{P}^{{l}},\mathbf{C}^{{l}} \right)- \sum_{m\in \mathcal{M}} \sum_{n\in \mathcal{N}}  \bar{R}_{k,m,n}^{\mathrm{appr}}\left(\mathbf{X}; \mathbf{X}^{{l}}, \mathbf{P}^{{l}},\mathbf{C}^{{l}} \right).
\end{aligned}
\end{equation}

We then obtain an approximation function of $Z(\mathbf{X}, \mathbf{P}^{{l}},\mathbf{C}^{{l}})$ as
\begin{equation}
\begin{aligned}
	&Z^\mathrm{Pos}\left(\mathbf{X}; \mathbf{X}^{{l}}, \mathbf{P}^{{l}},\mathbf{C}^{{l}} \right)=\min\limits_{k\in \mathcal{K}} 
	R_{k}^{\mathrm{Pos}}\left(\mathbf{X}; \mathbf{X}^{{l}}, \mathbf{P}^{{l}},\mathbf{C}^{{l}} \right) + \rho(\mathbf{\Lambda}^L, \mathbf{C}^{l})
\end{aligned}
\end{equation}
From \eqref{eq_R_kmn_appro1} and \eqref{eq_R_kmn_appro2}, it is not hard to verify that $Z^\mathrm{Pos}\left(\mathbf{X}; \mathbf{X}^{{l}}, \mathbf{P}^{{l}},\mathbf{C}^{{l}} \right)$ is concave in terms of $\mathbf{X}$.
Also, \eqref{c:value} and \eqref{c:gradient} hold because the functions in \eqref{eq_R_kmn_appro1} and \eqref{eq_R_kmn_appro2} are from the first-order Taylor approximations of that in $Z(\mathbf{X}, \mathbf{P}^l, \mathbf{C}^l)$ 	in terms of $\mathbf{X}$. 

Finally, the left-hand-side (LHS)  of constraint~\eqref{c:secure_dis} is lower-bounded by its first-order Taylor expansion w.r.t. $\|\mathbf{x}_m -\mathbf{x}_j\|$, i.e.,  
$
\|\mathbf{x}_m -\mathbf{x}_j\|^2 \geq 2\|\mathbf{x}_m^{l} -\mathbf{x}_j^{l}\|^\mathrm{T} \|\mathbf{x}_m-\mathbf{x}_j\|-\|\mathbf{x}_m^{l} -\mathbf{x}_j^{l}\|^2.
$
Then, constraint~\eqref{c:secure_dis} can be relaxed as the following convex constraint 
\begin{equation}\label{c:secure_dis_2}
\begin{aligned}
	&2\|\mathbf{x}_m^{l} -\mathbf{x}_j^{l}\|^\mathrm{T} \|\mathbf{x}_m -\mathbf{x}_j\|-\|\mathbf{x}_m^{l} -\mathbf{x}_j^{l}\|^2 \geq d_\mathrm{min}^2,  m,j\in \mathcal{M}, m\neq j.
\end{aligned}
\end{equation}

\subsubsection{Solution for Relaxed Problem}
Given a local point $\mathbf{X}^{l}$ in the $(l+1)$-th iteration, by replacing  $Z\left(\mathbf{X}, \mathbf{P}^{{l}},\mathbf{C}^{{l}} \right)$ with $Z^\mathrm{Pos}\left(\mathbf{X}; \mathbf{X}^{{l}}, \mathbf{P}^{{l}},\mathbf{C}^{{l}} \right)$ and replacing constraint~\eqref{c:secure_dis} with \eqref{c:secure_dis_2}, problem~\eqref{eq_problem_positioning} is relaxed as
\begin{align}
\max\limits_{\mathbf{X}}~~ &  Z^\mathrm{Pos}\left(\mathbf{X}; \mathbf{X}^{{l}}, \mathbf{P}^{{l}},\mathbf{C}^{{l}} \right)  \label{eq_problem_positioning_approx}\\
\mbox{s.t.}~~  
&\eqref{c:region}, \eqref{c:secure_dis_2}. \notag
\end{align}
Problem \eqref{eq_problem_positioning_approx} is to maximize a concave objective function subject to convex constraints on $\mathbf{X}$. Therefore, it is a convex problem and can be readily solved by standard convex program
solvers, such as CVXPY~\cite{diamond2016cvxpy}. 
We denote the optimal solution for problem~\eqref{eq_problem_positioning_approx} as $\tilde{\mathbf{X}}^{l\star}=\{ \tilde{\mathbf{x}}_m^{{l\star}} \}$.

\subsubsection{UAV Positioning Update}
Since $\tilde{\mathbf{X}}^{l\star}$ is an optimal solution for problem~\eqref{eq_problem_positioning_approx}, we have 
\begin{equation}\label{eq_pos_update_Reason}
\begin{aligned}
	0 &\overset{(e)}{\leq} Z^\mathrm{Pos}\left(\tilde{\mathbf{X}}^{l\star}; \mathbf{X}^{{l}}, \mathbf{P}^{{l}},\mathbf{C}^{{l}} \right) - Z^\mathrm{Pos}\left(\mathbf{X}^{{l}}; \mathbf{X}^{{l}}, \mathbf{P}^{{l}},\mathbf{C}^{{l}} \right) \\
	& \overset{(f)}{\leq}  (\tilde{\mathbf{X}}^{l\star}-\mathbf{X}^{{l}})^\mathrm{T}	\nabla_{\mathbf{X}} Z^\mathrm{Pos} \left(\mathbf{X}^{{l}}; \mathbf{X}^{{l}}, \mathbf{P}^{{l}},\mathbf{C}^{{l}} \right)    \overset{(g)}{=}  (\tilde{\mathbf{X}}^{l\star}-\mathbf{X}^{{l}})^\mathrm{T}	\nabla_{\mathbf{X}}   Z\left(\mathbf{X}^{{l}}, \mathbf{P}^{{l}},\mathbf{C}^{{l}} \right), 
\end{aligned}
\end{equation}
where $(e)$, $(f)$, and $(g)$ hold, respectively, due to  the optimality of $\tilde{\mathbf{X}}^{l\star}$, the concavity of $Z^\mathrm{Pos}\left(\mathbf{X}; \mathbf{X}^{{l}}, \mathbf{P}^{{l}},\mathbf{C}^{{l}} \right)$ w.r.t. $\mathbf{X}$, and the consistency conditions \eqref{c:value} and \eqref{c:gradient}. Therefore, $ \tilde{\mathbf{X}}^{l\star}-\mathbf{X}^{{l}}$ is an ascent direction of 
$Z\left(\mathbf{X}, \mathbf{P}^{{l}},\mathbf{C}^{{l}}\right)$
at $\mathbf{X}=\mathbf{X}^{l}$, along which
the objective value can be further increased compared with that at the current local point $\mathbf{X}^{l}$.

Given the ascent direction $\tilde{\mathbf{X}}^{l\star}-\mathbf{X}^{{l}}$, 
the UAV positioning variables are updated  as 
\begin{align}\label{eq_update_Position}
	\mathbf{X}^{l+1}=\mathbf{X}^{{l}}+ \gamma_1^l (\tilde{\mathbf{X}}^{l\star}-\mathbf{X}^{{l}}),
\end{align}
where $\gamma_1^l$ is the stepsize that needs to be selected properly. Backtracking line search can be adopted to choose $\gamma_1^l$ to efficiently increase $Z\left(\mathbf{X}, \mathbf{P}^{{l}},\mathbf{C}^{{l}}\right)$ 
and yield fast convergence~\cite{boyd2004convex}. Given constants  $\zeta, \tau  \in (0,1)$, $\gamma_1^l$ is set to be $\gamma_1^l=\zeta^{t_l}$ with $t_l$ being the smallest nonnegative integer that ensures     
constraint \eqref{c:secure_dis}, i.e., $\|\mathbf{x}_m^{l+1} -\mathbf{x}_j^{l+1}\|^2  \geq d_\mathrm{min}^2, m,j\in \mathcal{M}, m\neq j$, as well as the following inequality:\footnote{Constant $\tau$ can be interpreted as the fraction of the increment in $Z(\mathbf{X}, \mathbf{P}^l, \mathbf{C}^l)$ predicted by linear extrapolation that we will accept. 
Starting from $\gamma_1^l=\zeta^{0}=1$ and gradually decreasing $\gamma_1^l$, there always exists a $\gamma_1^l$ (at least we have $\gamma_1^l=\zeta^\infty=0$) satisfying both \eqref{c:secure_dis} and \eqref{con_stepsize}. For numerical evaluation, if $\gamma_1^l$ falls below a small positive threshold, we set $\gamma_1^l=0$, and update UAV positioning variables as $\mathbf{X}^{l+1}=\mathbf{X}^{{l}}$. }
\begin{equation}\label{con_stepsize}
	\begin{aligned}
		& Z\left(\mathbf{X}^{{l+1}}, \mathbf{P}^{{l}},\mathbf{C}^{{l}} \right)- Z\left(\mathbf{X}^{{l}}, \mathbf{P}^{{l}},\mathbf{C}^{{l}} \right)\geq \tau \gamma_1^l  (\tilde{\mathbf{X}}^{l\star}-\mathbf{X}^{{l}})^\mathrm{T}  \nabla_{\mathbf{X}} Z\left(\mathbf{X}^{{l}}, \mathbf{P}^{{l}},\mathbf{C}^{{l}} \right),
	\end{aligned}
\end{equation}
with
\begin{equation}\notag 
	\begin{aligned}
		&(\tilde{\mathbf{X}}^{l\star}-\mathbf{X}^{{l}})^\mathrm{T}  \nabla_{\mathbf{X}} Z\left(\mathbf{X}^{{l}}, \mathbf{P}^{{l}},\mathbf{C}^{{l}} \right) = \sum_{m\in \mathcal{M}} \sum_{n\in \mathcal{N}}  
		(\tilde{\mathbf{X}}^{l\star}-\mathbf{X}^{{l}})^\mathrm{T} 
		\nabla_{\mathbf{X}} R_{k',m,n}\left(\mathbf{X}^{{l}}, \mathbf{P}^{{l}},\mathbf{C}^{{l}} \right) \\
		&= \sum_{m\in \mathcal{M}} \sum_{n\in \mathcal{N}}  
		\left( 
		B_{k',m,n}^{{l}} \sum_{j \in \mathcal{M}}  \frac{p_{j,n}^{{l}}}{\sigma^2} \left( \tilde{\mathbf{x}}_j^{{l\star}}-\mathbf{x}_j^{{l}}\right)^\mathrm{T}  \nabla_{\mathbf{x}_j} g_{k'}\left(\mathbf{x}_j^{{l}} \right) 
		\right. \\
		&\left.
		~~~~-B_{k',-m,n}^{{l}} \sum_{j \in \mathcal{M}\setminus \{m\}}  \frac{p_{j,n}^{{l}}}{\sigma^2} \left( \tilde{\mathbf{x}}_j^{{l\star}}-\mathbf{x}_j^{{l}}\right)^\mathrm{T} \nabla_{\mathbf{x}_j} g_{k'}\left(\mathbf{x}_j^{{l}} \right) 
		\right),
	\end{aligned}
\end{equation}
and $k'= \arg \min\limits_{k} R_{k}\left(\mathbf{X}^{{l}}, \mathbf{P}^{{l}},\mathbf{C}^{{l}} \right) $ such that $Z\left(\mathbf{X}^{{l}}, \mathbf{P}^{{l}},\mathbf{C}^{{l}} \right)=R_{k'}\left(\mathbf{X}^{{l}}, \mathbf{P}^{{l}},\mathbf{C}^{{l}} \right) + \rho(\mathbf{\Lambda}^L, \mathbf{C}^{l})$.
The corresponding objective value of problem~\eqref{eq_problem_relaxed} is updated as  $Z(\mathbf{X}^{l+1},\mathbf{P}^{l},\mathbf{C}^{l})$.

\subsection{Resource Allocation}\label{sub_Sec_RA}
Given UAVs' positions $\mathbf{X}^{l+1}=\{\mathbf{x}_m^{{l+1}}\}$,
problem \eqref{eq_problem_MultiUAV}  is transformed to the following resource allocation problem
\begin{align}
\max\limits_{\mathbf{P}, \mathbf{C}}~ & Z(\mathbf{X}^{{l+1}}, \mathbf{P},\mathbf{C})           \label{eq_problem_ResourceAllocation}\\
\mbox{s.t.}~ 
&\eqref{c:orthogonal},\eqref{c:onetoone},  \eqref{c:power0},\eqref{c:power_max}, \eqref{c:c_kmn_continuous}.\notag
\end{align}
Problem~\eqref{eq_problem_ResourceAllocation} is a non-convex problem  because of the non-concave objective function. Similar to the method for optimizing UAV positioning, we aim to design a concave function approximation for $Z(\mathbf{X}^{{l+1}}, \mathbf{P},\mathbf{C})$, solve the relaxed counterpart of problem~\eqref{eq_problem_ResourceAllocation}, and update the resource allocation variables.
\subsubsection{Construction of Approximation Functions}
We aim to design a surrogate function $Z^\mathrm{RA}\left(\mathbf{P},\mathbf{C};\right.$ $\left. \mathbf{X}^{{l+1}}, \mathbf{P}^{{l}},\mathbf{C}^{{l}} \right)$, which is a concave function approximation of  $Z(\mathbf{X}^{{l+1}}, \mathbf{P},\mathbf{C})$ at local point $(\mathbf{P}^{{l}},\mathbf{C}^{{l}})$. Besides, the following conditions should be satisfied
\begin{align}
	&Z^\mathrm{RA}\left(\mathbf{P}^{{l}},\mathbf{C}^{{l}}; \mathbf{X}^{{l+1}}, \mathbf{P}^{{l}},\mathbf{C}^{{l}} \right) =Z(\mathbf{X}^{{l+1}}, \mathbf{P}^{l},\mathbf{C}^{l}), \label{c:RA_value} \\
	&\nabla_{\mathbf{P}} Z^\mathrm{RA}\left(\mathbf{P}^{{l}},\mathbf{C}^{{l}}; \mathbf{X}^{{l+1}}, \mathbf{P}^{{l}},\mathbf{C}^{{l}} \right) = \nabla_{\mathbf{P}} Z\left(\mathbf{X}^{{l+1}}, \mathbf{P}^{{l}}, \mathbf{C}^{{l}} \right), \label{c:RA_gradient1}\\
	&\nabla_{\mathbf{C}} Z^\mathrm{RA}\left(\mathbf{P}^{{l}},\mathbf{C}^{{l}}; \mathbf{X}^{{l+1}}, \mathbf{P}^{{l}},\mathbf{C}^{{l}} \right) = \nabla_{\mathbf{C}} Z\left(\mathbf{X}^{{l+1}}, \mathbf{P}^{{l}}, \mathbf{C}^{{l}} \right). \label{c:RA_gradient2}
\end{align}

To this end, we adopt the approximation of $R_{k,m,n}\left(\mathbf{X}^{{l+1}}, \mathbf{P}, \mathbf{C} \right)$ in the $(l+1)$-th iteration as	
\begin{equation}\label{eq_R_kmn_appro_RA}
\begin{aligned}
	&	R_{k,m,n}\left(\mathbf{X}^{{l+1}}, \mathbf{P}, \mathbf{C} \right)\approx R_{k,m,n}\left(\mathbf{X}^{{l+1}}, \mathbf{P}, \mathbf{C}^l \right)+ R_{k,m,n}\left(\mathbf{X}^{{l+1}}, \mathbf{P}^l, \mathbf{C} \right) - R_{k,m,n}\left(\mathbf{X}^{{l+1}}, \mathbf{P}^l, \mathbf{C}^l \right).
\end{aligned}
\end{equation}

Note that $R_{k,m,n}\left(\mathbf{X}^{{l+1}}, \mathbf{P}, \mathbf{C}^l \right)=\hat{R}_{k,m,n}\left(\mathbf{X}^{{l+1}}, \mathbf{P}, \mathbf{C}^l \right) - \bar{R}_{k,m,n}\left(\mathbf{X}^{{l+1}}, \mathbf{P}, \mathbf{C}^l \right)$
is the difference of two concave functions w.r.t. $\mathbf{P}$. Since any concave function is globally upper-bounded by its first-order Taylor expansion at any point, we have
\begin{equation}\label{eq_R_kmn2LB_PA}
\begin{aligned}
	&\bar{R}_{k,m,n}\left(\mathbf{X}^{{l+1}}, \mathbf{P}, \mathbf{C}^{l} \right)=c_{k,m,n}^{l} \log \left(1+\sum_{j \in \mathcal{M}\setminus \{m\}} \frac{g_k(\mathbf{x}_j^{{l+1}})}{\sigma^2} p_{j,n} \right)\\
	&\leq B_{k,-m,n}^{'{l}}\sum_{j \in \mathcal{M} /\{m\} }  \frac{g_k(\mathbf{x}_j^{{l+1}})}{\sigma^2}  \left( p_{j,n} - p_{j,n}^{{l}} \right) + \bar{R}_{k,m,n}\left(\mathbf{X}^{{l+1}}, \mathbf{P}^{l}, \mathbf{C}^{l} \right)\\
	&\triangleq \bar{R}_{k,m,n}^{\mathrm{ub}}\left(\mathbf{P}; \mathbf{X}^{{l+1}}, \mathbf{P}^{{l}}, \mathbf{C}^{{l}} \right),
\end{aligned}
\end{equation}
with $B_{k,-m,n}^{'{l}}=c_{k,m,n}^{{l}}/\left(1+\sum_{j \in \mathcal{M}\setminus \{m\}}  \frac{p_{j,n}^{{l}}}{\sigma^2} g_k(\mathbf{x}_j^{{l+1}})\right)$. 
Therefore, $R_{k,m,n}\left(\mathbf{X}^{{l+1}}, \mathbf{P}, \mathbf{C}^{l} \right)$ is lower-bounded by
\begin{equation}\label{eq_R_kmn_part_PA}
\begin{aligned}
	&R_{k,m,n}\left(\mathbf{X}^{{l+1}}, \mathbf{P}, \mathbf{C}^{l} \right) \geq  \hat{R}_{k,m,n}\left(\mathbf{X}^{{l+1}}, \mathbf{P}, \mathbf{C}^{l} \right)-
	\bar{R}_{k,m,n}^{\mathrm{ub}}\left(\mathbf{P}; \mathbf{X}^{{l+1}}, \mathbf{P}^{{l}}, \mathbf{C}^{{l}} \right).
\end{aligned}
\end{equation}

By substituting \eqref{eq_R_kmn_part_PA} into \eqref{eq_R_kmn_appro_RA}, $	R_{k,m,n}\left(\mathbf{X}^{{l+1}}, \mathbf{P}, \mathbf{C} \right)$ is further approximated as 
\begin{equation}\label{eq_R_kmn_appro_RA2}
\begin{aligned}
	R_{k,m,n}\left(\mathbf{X}^{{l+1}}, \mathbf{P}, \mathbf{C} \right)&\approx \hat{R}_{k,m,n}\left(\mathbf{X}^{{l+1}}, \mathbf{P}, \mathbf{C}^{l} \right)-
	\bar{R}_{k,m,n}^{\mathrm{ub}}\left(\mathbf{P}; \mathbf{X}^{{l+1}}, \mathbf{P}^{{l}}, \mathbf{C}^{{l}} \right) \\&~~~+ R_{k,m,n}\left(\mathbf{X}^{{l+1}}, \mathbf{P}^{l}, \mathbf{C} \right)- R_{k,m,n}\left(\mathbf{X}^{{l+1}}, \mathbf{P}^{l}, \mathbf{C}^{l} \right) \\& \triangleq R_{k,m,n}^\mathrm{appr} \left( \mathbf{P}, \mathbf{C};\mathbf{X}^{{l+1}}, \mathbf{P}^{l}, \mathbf{C}^{l} \right).
\end{aligned}
\end{equation}

In addition, since $c_{k,m,n}^2\geq 2c_{k,m,n}^l c_{k,m,n}-(c_{k,m,n}^l)^2$ for a given local point $c_{k,m,n}^l$, we have the following lower bound on $\rho(\mathbf{\Lambda}^L, \mathbf{C})$:
\begin{equation}\label{eq_Penalty_Appro}
\begin{aligned}
	\rho(\mathbf{\Lambda}^L, \mathbf{C}) &= -\sum_{k \in \mathcal{K}}\sum_{m \in \mathcal{M}}\sum_{n \in \mathcal{N}} \lambda_{k,m,n}^L  c_{k,m,n}(1-c_{k,m,n}) \\&
	\geq \sum_{k \in \mathcal{K}}\sum_{m \in \mathcal{M}}\sum_{n \in \mathcal{N}} \lambda_{k,m,n}^L \left( \left(2c_{k,m,n}^{l}-1\right)c_{k,m,n}-\left(c_{k,m,n}^{l} \right)^2 \right)\\& \triangleq \rho^\mathrm{lb}(\mathbf{C};\mathbf{\Lambda}^L, \mathbf{C}^{l}).
\end{aligned}
\end{equation}

According to \eqref{eq_R_kmn_appro_RA2} and \eqref{eq_Penalty_Appro}, we obtain an approximation function of $Z\left(\mathbf{X}^{{l+1}}, \mathbf{P}, \mathbf{C} \right)$ as 
\begin{equation}\label{eq_R_kmn_appro_RA_Final}
\begin{aligned}
	&Z^\mathrm{RA}\left(\mathbf{P},\mathbf{C}; \mathbf{X}^{{l+1}}, \mathbf{P}^{{l}},\mathbf{C}^{{l}} \right)= \min\limits_{ k\in \mathcal{K}} \left(
	\sum_{m\in \mathcal{M}} \sum_{n\in \mathcal{N}}  R_{k,m,n}^\mathrm{appr} \left( \mathbf{P}, \mathbf{C};\mathbf{X}^{{l+1}}, \mathbf{P}^{l}, \mathbf{C}^{l} \right)  
	\right)+\rho^\mathrm{lb}(\mathbf{C};\mathbf{\Lambda}^L, \mathbf{C}^{l}).
\end{aligned}
\end{equation}
It can be verified that conditions \eqref{c:RA_value}, \eqref{c:RA_gradient1}, and \eqref{c:RA_gradient2} are satisfied.

\subsubsection{Solution for Relaxed Problem}
Problem~\eqref{eq_problem_ResourceAllocation} can be relaxed as the following convex problem
\begin{align}
\max\limits_{\mathbf{P}, \mathbf{C}}~ & Z^\mathrm{RA}\left(\mathbf{P},\mathbf{C}; \mathbf{X}^{{l+1}}, \mathbf{P}^{{l}},\mathbf{C}^{{l}} \right)          \label{eq_problem_ResourceAllocation_Approx}\\
\mbox{s.t.}~
&\eqref{c:orthogonal},\eqref{c:onetoone},  \eqref{c:power0},\eqref{c:power_max}, \eqref{c:c_kmn_continuous},\notag
\end{align}
which  can be readily solved by standard convex program
solvers. We denote the optimal solution for problem~\eqref{eq_problem_ResourceAllocation_Approx} as $(\tilde{\mathbf{P}}^{l\star}=\{\tilde{p}_{m,n}^{l\star}\}, \tilde{\mathbf{C}}^{l\star}=\{\tilde{c}_{k,m,n}^{l\star}\})$, which is an ascent direction of 
$Z\left(\mathbf{X}^{l+1}, \mathbf{P},\mathbf{C}\right)$
at $(\mathbf{P}^{l}, \mathbf{C}^{l})$.

\subsubsection{Resource Allocation Update}
We update the resource allocation variables as 
\begin{align}\label{eq_update_Resource}
\left\{\begin{matrix}
	\mathbf{P}^{l+1}= \mathbf{P}^{l} + \gamma_2^{l} (\tilde{\mathbf{P}}^{l\star}-\mathbf{P}^{l}),
	\\
	\mathbf{C}^{l+1}=\mathbf{C}^{l} + \gamma_2^{l} (\tilde{\mathbf{C}}^{l\star}-\mathbf{C}^{l}),
\end{matrix}\right.
\end{align}
where $\gamma_2^{l}$ is the stepsize determined by adopting the backtracking line search, with the following condition satisfied:
\begin{equation}\label{con_stepsizeRA}
\begin{aligned}
	&  Z\left(\mathbf{X}^{{l+1}}, \mathbf{P}^{{l+1}}, \mathbf{C}^{{l+1}} \right) - Z\left(\mathbf{X}^{{l+1}}, \mathbf{P}^{{l}}, \mathbf{C}^{{l}} \right)\geq \tau \gamma_2^l 
	\begin{bmatrix}
		(\tilde{\mathbf{P}}^{l\star}-\mathbf{P}^{{l}} )^\mathrm{T}	\\ (\tilde{\mathbf{C}}^{l\star}-\mathbf{C}^{{l}})^\mathrm{T}
	\end{bmatrix}^\mathrm{T}
	\begin{bmatrix}
		\nabla_{\mathbf{P}} Z\left(\mathbf{X}^{{l+1}}, \mathbf{P}^{{l}}, \mathbf{C}^{{l}} \right)	\\ \nabla_{\mathbf{C}} Z\left(\mathbf{X}^{{l+1}}, \mathbf{P}^{{l}}, \mathbf{C}^{{l}} \right)
	\end{bmatrix},
\end{aligned}
\end{equation}
with
\begin{equation}\notag
\begin{aligned}
	&(\tilde{\mathbf{P}}^{l\star}-\mathbf{P}^{{l}})^\mathrm{T}
	\nabla_{\mathbf{P}} Z\left(\mathbf{X}^{{l+1}}, \mathbf{P}^{{l}}, \mathbf{C}^{{l}} \right)= \sum_{m\in \mathcal{M}} \sum_{n\in \mathcal{N}}  
	(\tilde{\mathbf{P}}^{l\star}-\mathbf{P}^{{l}})^\mathrm{T} 
	\nabla_{\mathbf{P}} R_{k',m,n}\left(\mathbf{X}^{{l+1}}, \mathbf{P}^{{l}},\mathbf{C}^{{l}} \right) \\
	&= \sum_{m\in \mathcal{M}} \sum_{n\in \mathcal{N}}  
	\left( 
	B_{k',m,n}^{'{l}}\sum_{j \in \mathcal{M}  }  \frac{g_{k'}(\mathbf{x}_j^{{l+1}})}{\sigma^2}  \left( \tilde{p}_{j,n}^{{l\star}} - p_{j,n}^{{l}} \right)
-B_{k',-m,n}^{'{l}}\sum_{j \in \mathcal{M} /\{m\} }  \frac{g_{k'}(\mathbf{x}_j^{{l+1}})}{\sigma^2}  \left( \tilde{p}_{j,n}^{{l\star}} - p_{j,n}^{{l}} \right)
	\right),
\end{aligned}
\end{equation}
\begin{equation}\notag
\begin{aligned}
	(\tilde{\mathbf{C}}^{l\star}-\mathbf{C}^{{l}})^\mathrm{T} \nabla_{\mathbf{C}} Z\left(\mathbf{X}^{{l+1}}, \mathbf{P}^{{l}}, \mathbf{C}^{{l}} \right)=&\sum_{m\in \mathcal{M}} \sum_{n\in \mathcal{N}} (\tilde{c}_{k',m,n}^{l\star}-c_{k',m,n}^{l})
	\frac{R_{k',m,n}\left(\mathbf{X}^{{l+1}}, \mathbf{P}^{{l}}, \mathbf{C}^{{l}} \right)}{c_{k',m,n}^{l}}
	\\
	& -  \sum_{m\in \mathcal{M}} \sum_{n\in \mathcal{N}}\sum\limits_{k \in \mathcal{K}} (\tilde{c}_{k,m,n}^{l\star}-c_{k,m,n}^{l}) \lambda_{k,m,n}^{L}  (1-2c_{k,m,n}^{l}) 
	,
\end{aligned}
\end{equation}
and $k'= \arg \min\limits_{k} R_{k}\left(\mathbf{X}^{{l+1}}, \mathbf{P}^{{l}},\mathbf{C}^{{l}} \right) $, $B_{k',m,n}^{'{l}}=c_{k',m,n}^{{l}}/\left(1+\sum_{j \in \mathcal{M}}  \frac{p_{j,n}^{{l}}}{\sigma^2} g_{k'} \left(\mathbf{x}_j^{{l+1}} \right)\right)$. 
The corresponding objective value of problem~\eqref{eq_problem_relaxed} is updated as $Z(\mathbf{X}^{l+1},\mathbf{P}^{l+1},\mathbf{C}^{l+1})$.


\subsection{Penalty Multiplier Update}\label{sub_Sec_Multiplier}
For the  $(L+1)$-th outer-loop with given $\Lambda^{L}$, we have obtained a suboptimal solution for problem~\eqref{eq_problem_relaxed} by inner-loop iterations (as described in Sections~\ref{sub_Sec_Positioning} and~\ref{sub_Sec_RA}), denoted by $\bar{\mathbf{X}}^{L+1}=\{\bar{\mathbf{x}}_m^{L+1}\}$, $ \bar{\mathbf{P}}^{L+1}=\{\bar{p}_{m,n}^{L+1}\}$, and $\bar{\mathbf{C}}^{L+1}=\{\bar{c}_{k,m,n}^{L+1}\}$.
To decrease the violation of constraint \eqref{c:c_kmn_continuous2}, the multipliers in $\mathbf{\Lambda}$ are updated by using the following strategy~\cite{Fisher1981The}
\begin{align}\label{update_multiplier}
\lambda_{k,m,n}^{L+1}=\lambda_{k,m,n}^{L}+ \gamma^L \bar{c}_{k,m,n}^{L+1}(1-\bar{c}_{k,m,n}^{L+1}), 
\end{align}
which basically optimizes a dual problem w.r.t. $\mathbf{\Lambda}$ via the gradient method~\cite{bertsekas1997nonlinear}. $\gamma^L$ is the stepsize given by
\begin{align}
\gamma^L=\frac{\mu^L}{\sum\limits_{k \in \mathcal{K}}\sum\limits_{m \in \mathcal{M}}\sum\limits_{n \in \mathcal{N}} \left( \bar{c}_{k,m,n}^{L+1}(1-\bar{c}_{k,m,n}^{L+1}) \right)^2  },
\end{align}
where $\mu^L$ is an adaption parameter which is set as $\mu^0=2$ and $\mu^L \leftarrow 2\mu^L$ for the case when the maximum constraint violation $\max\limits_{k ,m,n} \bar{c}_{k,m,n}^{L}(1-\bar{c}_{k,m,n}^{L})$ does not decrease in the $L$-th iteration. 

It can be seen that \eqref{update_multiplier} guarantees an increasing update for $\lambda_{k,m,n}$. By gradually increasing the value of $\lambda_{k,m,n}$, the violation of constraint~\eqref{c:c_kmn_continuous2} can decrease. For sufficiently large $\lambda_{k,m,n}$, constraint~\eqref{c:c_kmn_continuous2} is strictly satisfied such that  the obtained solution for problem~\eqref{eq_problem_relaxed} is also a feasible solution for  the original problem~\eqref{eq_problem_MultiUAV}.

\subsection{Overall Solution}\label{sub_Sec_Overall}
Hereto, we are ready to show the overall PDLIO  algorithm for solving the joint positioning and resource allocation problem~\eqref{eq_problem_MultiUAV} for a multi-UAV OFDMA communication network aided by geographic information. As summarized in Algorithm~\ref{alg:overall_solution}, in line 1, we invoke~\cite[Algorithm~1]{yi2022joint3} to calculate the blocked regions $\{\mathcal{D}_{k,q},  k \in \mathcal{K}, q \in \mathcal{Q} \}$ for $K$ users caused by $Q$ buildings based on geographic information. Then, in lines 3-13, we employ the double-loop framework to iteratively solve the penalized problem~\eqref{eq_problem_relaxed} and update the multipliers $\mathbf{\Lambda}$. Lines 5-9 solve the penalized problem~\eqref{eq_problem_relaxed} with given $\mathbf{\Lambda}^{L}$, where the UAV positioning and resource allocation are optimized in an alternate manner. The inner-loop terminates when the increase of the objective value of problem~\eqref{eq_problem_relaxed} from one iteration to the next falls bellow a positive threshold $\epsilon_\mathrm{l}$.
The outer-loop terminates when the maximum constraint violation $\max\limits_{k,m,n} \bar{c}_{k,m,n}^{L}(1-\bar{c}_{k,m,n}^{L})$ falls below a positive threshold $\epsilon_\mathrm{L}$.

\begin{algorithm}[t] \small
\label{alg:overall_solution}
\caption{Joint positioning and resource allocation for multi-UAV OFDMA communication network.}
\begin{algorithmic}[1]
\REQUIRE ~The coordinates of the vertices of the buildings, $\{\mathbf{x}_k,  k\in \mathcal{K} \}$,  $P_\mathrm{max}$, $\sigma^2$,
$\alpha_1$, $\beta_1$, $\alpha_2$, $\beta_2$, $\eta$, $x_\mathrm{D}$, 
$y_\mathrm{D}$, $h_\mathrm{min}$, $d_\mathrm{min}$, $\zeta$,  $\tau$, $\epsilon_\mathrm{L}^{}$, $\epsilon_\mathrm{l}^{}$.
\STATE Calculate the blocked regions $\{\mathcal{D}_{k,q},  k \in \mathcal{K}, q \in \mathcal{Q} \}$ according to~\cite[Algorithm~1]{yi2022joint3}.
\STATE Set the iteration index of outer-loops as $L=0$, and initialize $\mathbf{\Lambda}^{0}$, $\bar{\mathbf{X}}^{0}$, $\bar{\mathbf{P}}^{0}$, $\bar{\mathbf{C}}^{0}$.		
\REPEAT
\STATE Set the iteration index of inner-loops as $l=0$, and initialize $\mathbf{X}^{0}\leftarrow \bar{\mathbf{X}}^{L}$, $\mathbf{P}^{0}\leftarrow\bar{\mathbf{P}}^{L}$, $\mathbf{C}^{0}\leftarrow\bar{\mathbf{C}}^{L}$.
\REPEAT
\STATE $\bullet$ Solve problem~\eqref{eq_problem_positioning_approx} to obtain the optimal solution $\tilde{\mathbf{X}}^{l\star}$, and update the UAV position $\mathbf{X}^{l+1}$ according to~\eqref{eq_update_Position}.
\STATE $\bullet$ Solve problem~\eqref{eq_problem_ResourceAllocation_Approx} to obtain the optimal solution $(\tilde{\mathbf{P}}^{l\star}, \tilde{\mathbf{C}}^{l\star})$, and update the power allocation $\mathbf{P}^{l+1}$ and association $\mathbf{C}^{l+1}$ according to~\eqref{eq_update_Resource}.
\STATE $\bullet$ Update $l\leftarrow l+1$.
\UNTIL{The increase of objective value falls bellow $\epsilon_\mathrm{l}$.}	
\STATE{Update $\bar{\mathbf{X}}^{L+1} \leftarrow \mathbf{X}^{l}$, $\bar{\mathbf{P}}^{L+1} \leftarrow \mathbf{P}^{l}$, $\bar{\mathbf{C}}^{L+1} \leftarrow \mathbf{C}^{l}$. }
\STATE Update  multiplier $\mathbf{\Lambda}^{L+1}$ based on \eqref{update_multiplier}.
\STATE Update $L\leftarrow L+1$.
\UNTIL{The maximum constraint violation $\max\limits_{k ,m ,n} \bar{c}_{k,m,n}^{L}(1-\bar{c}_{k,m,n}^{L})$ falls below $\epsilon_\mathrm{L}$.}		
\ENSURE   $\bar{\mathbf{X}}^{L}$, $\bar{\mathbf{P}}^{L}$,   $\bar{\mathbf{C}}^{L}$, $ Z\left(\bar{\mathbf{X}}^{L}, \bar{\mathbf{P}}^{L}, \bar{\mathbf{C}}^{L} \right) $.
\end{algorithmic}
\end{algorithm}

For the proposed algorithm, 
the calculation of blocked regions  in line 1 entails a computational complexity of $\mathcal{O}\left( KQ \right)$. The computational complexity of lines 5-9 is dominated by solving the 3-D positioning sub-problem~\eqref{eq_problem_positioning_approx}  and  resource allocation sub-problem~\eqref{eq_problem_ResourceAllocation_Approx}, whose complexities are $\mathcal{O}\left( (3M)^{3.5} \right)$ and $\mathcal{O}\left( (KMN+MN)^{3.5} \right)$, respectively, by using the interior point method~\cite{boyd2004convex}. The complexity of updating penalty multipliers in line~11 is $\mathcal{O}\left( KMN \right)$. Therefore, the worst-case computational complexity of Algorithm~\ref{alg:overall_solution} is $\mathcal{O}\left(KQ+ L_\mathrm{out} L_\mathrm{in} ((3M)^{3.5}+ (KMN+MN)^{3.5} ) \right)$, where $L_\mathrm{out}$ and $L_\mathrm{in}$ denote the numbers of the outer-loop and inner-loop iterations, respectively.

The inner-loop solves the penalized problem~\eqref{eq_problem_relaxed} by using the BSCA technique.
Due to the use of backtracking line search, we have
\begin{equation}\label{converge_1}
\begin{aligned}
&Z(\mathbf{X}^{{l+1}}, \mathbf{P}^{{l}},\mathbf{C}^{{l}}) - Z(\mathbf{X}^{{l}}, \mathbf{P}^{{l}},\mathbf{C}^{{l}}) \geq \tau \gamma_1^{l} (\tilde{\mathbf{X}}^{l\star}-\mathbf{X}^{{l}})^\mathrm{T}	\nabla_{\mathbf{X}}  Z\left(\mathbf{X}^{{l}}, \mathbf{P}^{{l}},\mathbf{C}^{{l}} \right) \geq 0, 
\end{aligned}
\end{equation}
and 
\begin{equation}\label{converge_2}
\begin{aligned}
&  Z\left(\mathbf{X}^{{l+1}}, \mathbf{P}^{{l+1}}, \mathbf{C}^{{l+1}} \right) - Z\left(\mathbf{X}^{{l+1}}, \mathbf{P}^{{l}}, \mathbf{C}^{{l}} \right)\geq \tau \gamma_2^l 
\begin{bmatrix}
	(\tilde{\mathbf{P}}^{l\star}-\mathbf{P}^{{l}} )^\mathrm{T}	\\ (\tilde{\mathbf{C}}^{l\star}-\mathbf{C}^{{l}})^\mathrm{T}
\end{bmatrix}^\mathrm{T}
\begin{bmatrix}
	\nabla_{\mathbf{P}} Z\left(\mathbf{X}^{{l+1}}, \mathbf{P}^{{l}}, \mathbf{C}^{{l}} \right)	\\ \nabla_{\mathbf{C}} Z\left(\mathbf{X}^{{l+1}}, \mathbf{P}^{{l}}, \mathbf{C}^{{l}} \right)
\end{bmatrix} \geq 0.
\end{aligned}
\end{equation}
Based on~\eqref{converge_1} and~\eqref{converge_2}, we have 
\begin{equation}\notag
\begin{aligned}
&  Z\left(\mathbf{X}^{{l+1}}, \mathbf{P}^{{l+1}}, \mathbf{C}^{{l+1}} \right) \geq Z\left(\mathbf{X}^{{l+1}}, \mathbf{P}^{{l}}, \mathbf{C}^{{l}} \right)\geq  Z\left(\mathbf{X}^{{l}}, \mathbf{P}^{{l}}, \mathbf{C}^{{l}} \right),
\end{aligned}
\end{equation}
which indicates that the objective value of problem~\eqref{eq_problem_relaxed} is non-decreasing over the iteration. Since the objective value is upper-bounded, it always converges to a finite value.
For the outer-loop, the increasing update for $\mathbf{\Lambda}$ leads to a sufficiently large multiplier which ensures that the solution for the penalized problem~\eqref{eq_problem_relaxed} converges to a feasible solution for the original problem in~\eqref{eq_problem_MultiUAV}. Therefore, Algorithm~\ref{alg:overall_solution} is guaranteed to obtain a suboptimal solution for problem~\eqref{eq_problem_MultiUAV}. The convergence  of Algorithm~\ref{alg:overall_solution} will be further evaluated by simulations in Section~\ref{SubSec_Simulation}.


\section{Performance Evaluation}\label{sec_simulation}
In this section, we provide simulation results to evaluate the performance of the proposed joint 3-D positioning and resource allocation scheme for multi-UAV communication networks aided by geographic information.

\subsection{Simulation Setup and Benchmark Schemes}
\begin{figure}[t]
\centering
\subfigure{\includegraphics[align=t, width=0.35\textwidth]{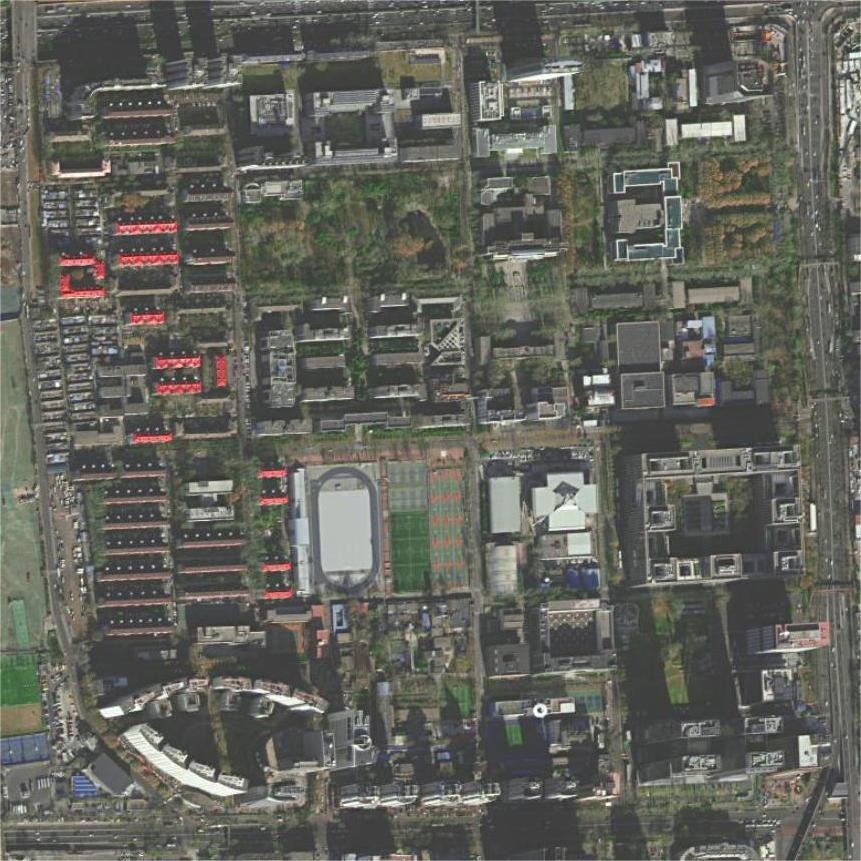}}
\subfigure{\includegraphics[align=t,width=0.48\textwidth]{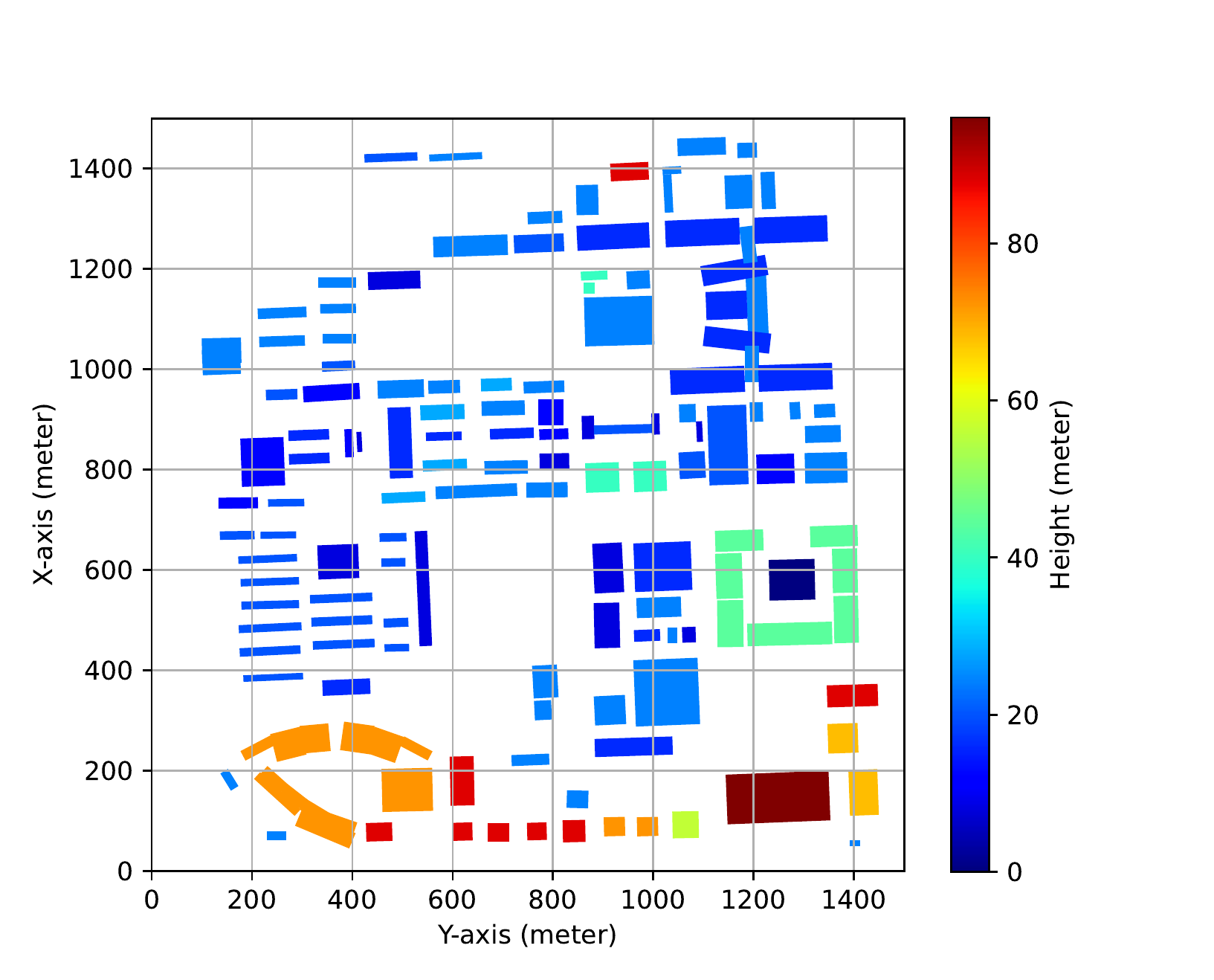}}
\caption{(Left) An orthoimagery of the $1, 500\times1, 500\text{ m}^2$ campus area at
Beihang University, Beijing, China. (Right) The corresponding elevation map of buildings.}
\label{fig:buildings}
\end{figure}

As shown in Fig.~\ref{fig:buildings}, we consider a dense urban area of campus at Beihang University with size $1, 500\times1, 500\text{ m}^2$, i.e., $x_\mathrm{D}=1, 500$, $y_\mathrm{D}=1, 500$. The geometries of buildings are obtained from the \emph{OpenStreetMap} database\footnote{https://www.openstreetmap.org/} and then processed into a number of cubes. In this area, the maximum building height is $96$ m. Therefore, the minimum flight altitude of UAVs, $h_\mathrm{min}$, is set to $100$ m  such that no collision will occur. The adopted simulation parameter settings are listed in Table~\ref{tab:para}~\cite{yi2022joint3,Al-Hourani2014modeli}, unless specified otherwise. The users are randomly generated on the ground, and each point in the simulation figures is the average performance over 500 user location realizations. 

\begin{table}[t]
\caption{Simulation Parameters}\label{tab:para}
\footnotesize
\begin{center}
\begin{tabular}{|c|l|c|}
	\hline
	\textbf{Parameter}                        &\multicolumn{1}{c|}{\textbf{Description}}                                       & \textbf{Value} \\
	\hline
	$P_{\mathrm{max}}$                                 & Maximum transmit power of each UAV                                   & 30 dBm \\
	\hline
	$\sigma^2$                           & Power of the noise at users                             & -107 dBm \\
	\hline
	$f_c$                       & Carrier frequency                                         & 5 GHz \\
	\hline
	$\alpha_{\mathrm{1}}$                    & Channel gain exponent for LoS path                          & 2 \\
	\hline
	$\alpha_{\mathrm{2}}$                   & Channel gain exponent for NLoS path                         & 3.3 \\
	\hline
	$\beta_{\mathrm{1}}$                  & \begin{tabular}[c]{@{}l@{}}Channel gain at the reference distance of   1 m for LoS path\end{tabular}                          & -46.43 dB \\
	\hline
	$\beta_{\mathrm{2}}$                   & \begin{tabular}[l]{@{}l@{}}Channel gain at the reference distance of  1 m for NLoS path \end{tabular}                        & -56.43 dB \\
	\hline
	$d_\mathrm{min}$                  & \begin{tabular}[c]{@{}l@{}}Minimum inter-UAV distance \end{tabular}                          & 25 m \\
	
	\hline
	$\eta$                   & \begin{tabular}[l]{@{}l@{}}Smooth parameter of the channel model \end{tabular}                        & 1000 \\
	\hline
	$\zeta$                                   & \begin{tabular}[l]{@{}l@{}}Scaling factor in backtracking line search \end{tabular}                              & 0.9 \\
	\hline
	$\tau$                                   & \begin{tabular}[l]{@{}l@{}}Coefficient in backtracking line search \end{tabular}                              & 0.01 \\
	\hline
	$\epsilon_{\mathrm{l}}^{}$                                 & \begin{tabular}[l]{@{}l@{}}Threshold for convergence of inner-loop \end{tabular}                  & $10^{-3}$ \\
	\hline
	$\epsilon_{\mathrm{L}}^{}$                                 & \begin{tabular}[l]{@{}l@{}}Threshold for convergence of outer-loop \end{tabular}                  & $10^{-4}$ \\
	\hline
	$\lambda_{k,m,n}^{0}$                                   & \begin{tabular}[l]{@{}l@{}}Initial value of the  multiplier \end{tabular}                              & $0.2K/(MN)$ \\
	\hline
\end{tabular}
\end{center}
\end{table}

The initial state of the system, i.e., $\{ \bar{\mathbf{X}}^{0},\bar{\mathbf{P}}^{0},\bar{\mathbf{C}}^{0}\}$, is determined via the following steps.
First, $M$ UAVs are deployed right above $M$ out of $K$ users at an altitude of 500 m, where the users that are the closest to the four endpoints of the area are sequentially selected.
Then, each user selects the one with the highest channel gain among the UAVs with idle subcarriers for UAV association, and occupies the least used idle subcarrier to minimize the interference. Finally, each UAV evenly allocates the transmit power on its occupied subcarriers.

We label the proposed method as ``Proposed", and define three benchmark schemes for performance comparison, namely ``Fixed Association", ``K-means Position", and ``No GoeInfo", respectively, explained as follows.  

\begin{itemize}
\item \emph{Fixed Association}: In this scheme, the association variables $\mathbf{C}$ are fixed as initial value $\bar{\mathbf{C}}^{0}$, while the UAV positioning and power allocation are jointly optimized following lines 5-9 in Algorithm~\ref{alg:overall_solution}. Note that there is no need to perform any outer-loop iteration since constraint~\eqref{c:c_kmn_continuous2} is always  satisfied with $\mathbf{C}=\bar{\mathbf{C}}^{0}$.  

\item \emph{K-means Position}: This scheme partitions the users into $M$ groups based on their horizontal coordinates by employing the K-means algorithm, and places the UAVs right above the cluster centers with an altitude of 500 m. Then, all the steps in Algorithm~\ref{alg:overall_solution} except line 6, are executed to optimize resource allocation. 

\item \emph{No GeoInfo}: This scheme assumes that geographic information is unavailable for the system. Therefore, the joint positioning and resource allocation is performed with the assumption of LoS A2G channels. The achievable rate is calculated according to the actual LoS/NLoS channels at the obtained positions of UAVs. 
\end{itemize}



\subsection{Simulation Results}\label{SubSec_Simulation}
\begin{figure}[t]
\centering
\subfigure[Initial state.]{\includegraphics[align=t, width=0.4\textwidth]{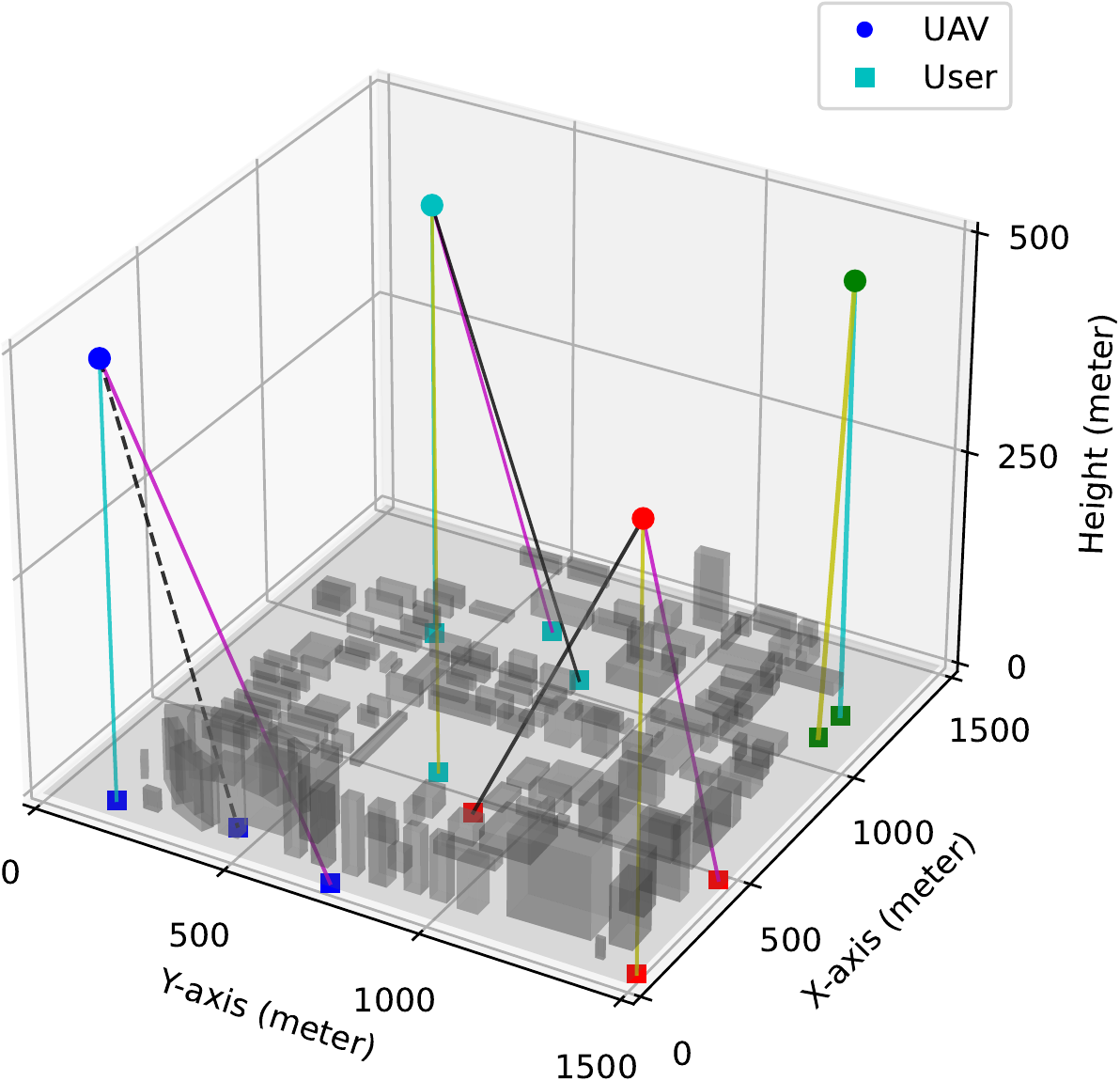}}
\subfigure[Final state.]{\includegraphics[align=t,width=0.4\textwidth]{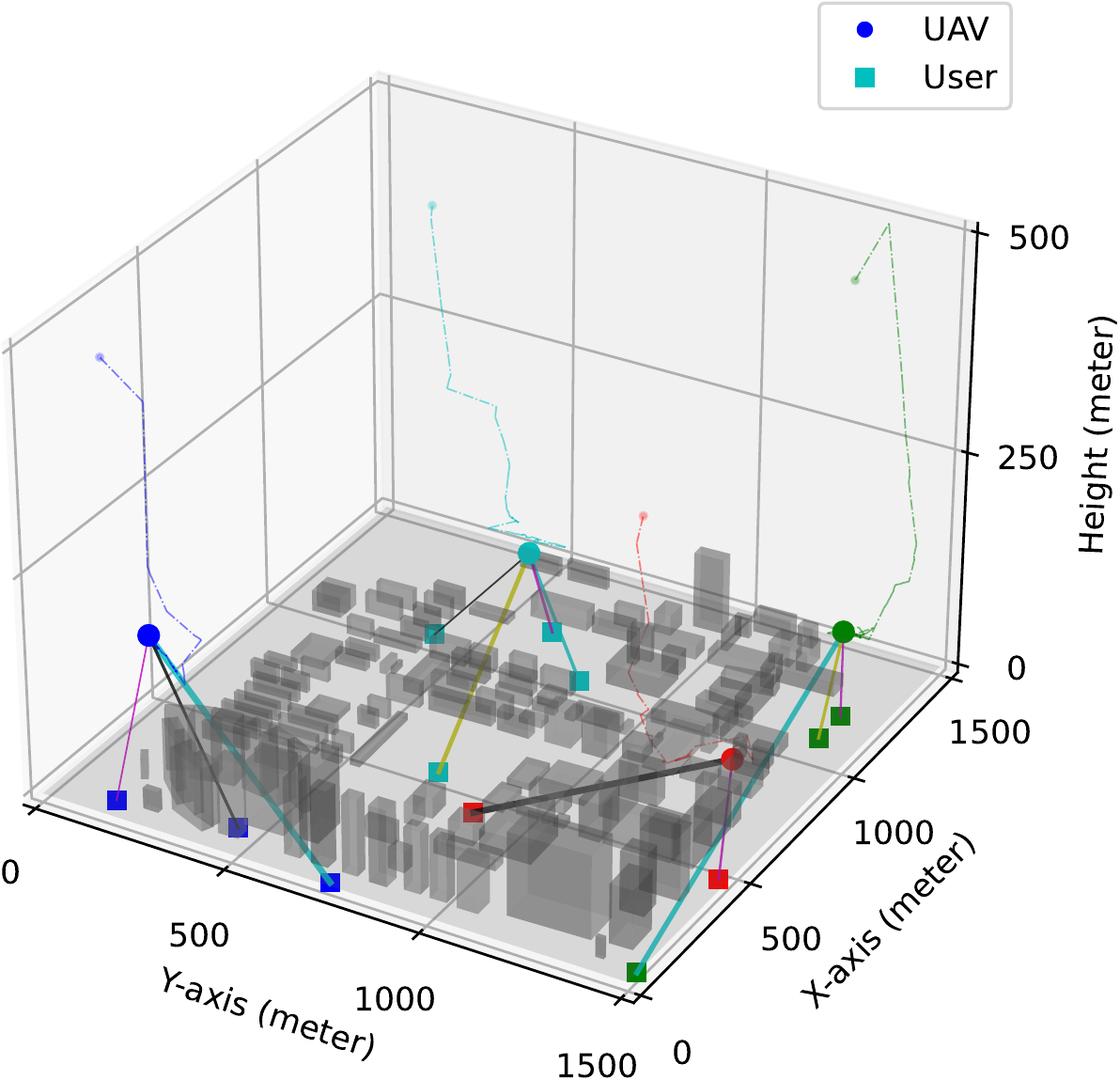}}
\caption{Demonstration of the proposed solution for the UAV positioning and resource allocation in Algorithm~\ref{alg:overall_solution}.}
\label{fig:optimization}
\end{figure}

First, we provide a demonstration of the proposed solution for the UAV positioning and resource allocation in Fig.~\ref{fig:optimization}. The users are marked by `$\blacksquare$', and the UAVs are marked by `$\bullet$'.
Each UAV and its associated users are marked with the same color, and are connected by a solid (dashed) line for LoS (NLoS) channel condition. Links with different colors represent that they occupy different subcarriers, and the linewidths are proportional to the allocated transmit powers. 
As shown in Fig.~\ref{fig:optimization}~(a), 4 UAVs are initially placed right above 4 out of 12 users at an altitude of 500 m.
Since there is one user who cannot establish LoS links with the UAVs, the minimum achievable rate  of the system is  only 0.042 bits/sec/Hz.   
The final state of the system by employing the proposed solution is shown in Fig.~\ref{fig:optimization}~(b), where the optimization process of the UAV positioning is demonstrated by a  dash-dotted line. As can be observed, the UAVs tend to decrease their altitudes and adjust their horizontal positions to decrease  path losses as well as get rid of  building blockages. Besides, users farther away from their associated UAVs are allocated with more power to compensate for higher path loss. User association and subcarrier allocation have also adapted  to coordinate mutual interference along with UAV positioning and power allocation. Compared to the initial state, the minimum achievable rate after optimization increases from 0.042 bits/sec/Hz to 5.344 bits/sec/Hz, which demonstrates the effectiveness of our proposed Algorithm~\ref{alg:overall_solution}.

\begin{figure}[t]
\centering
\subfigure[Maximum constraint violation versus the number of outer-loop iterations.]{\includegraphics[align=t, width=0.42\textwidth]{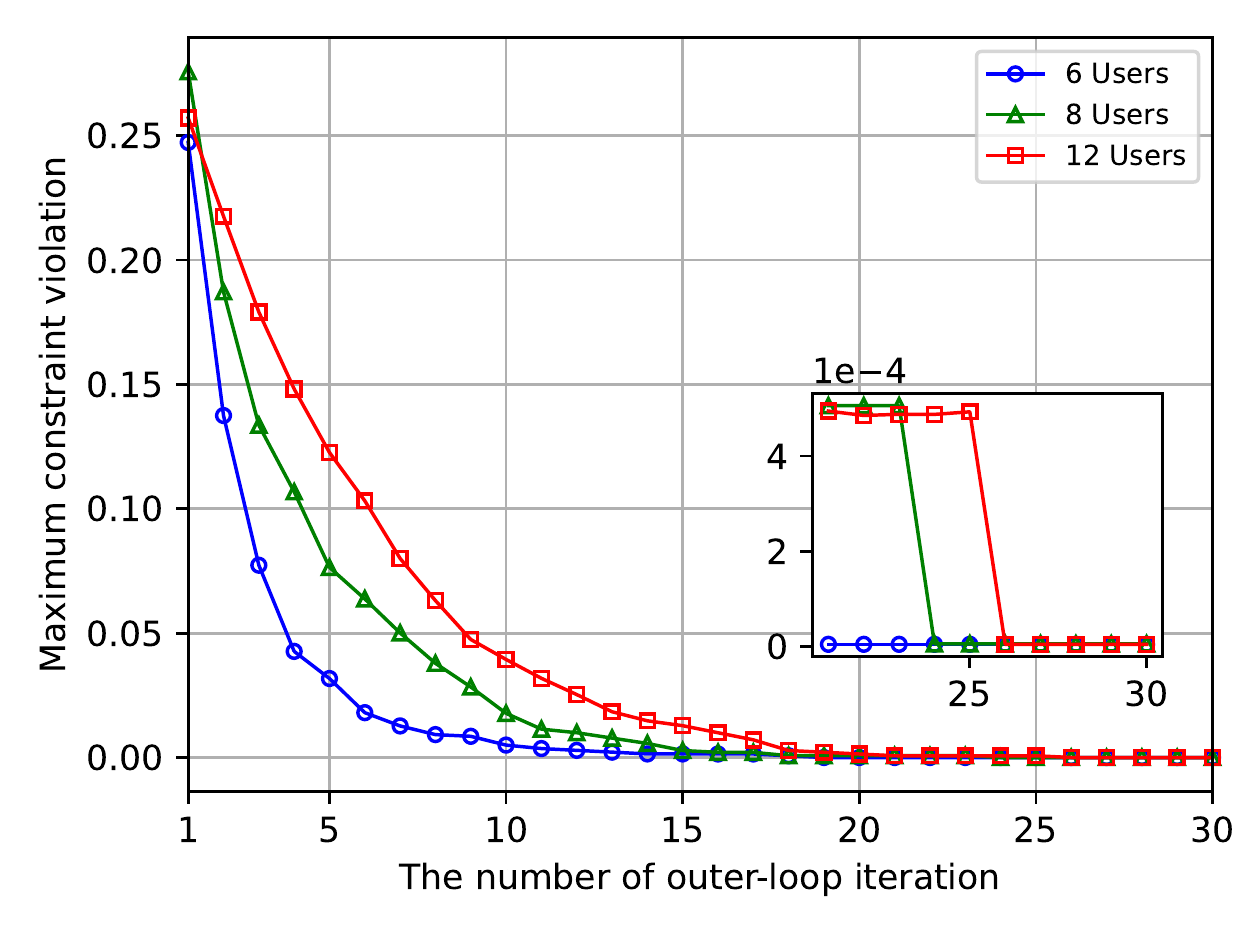}}
\subfigure[Objective value of problem~\eqref{eq_problem_relaxed} versus the number of inner-loop iterations.]{\includegraphics[align=t,width=0.42\textwidth]{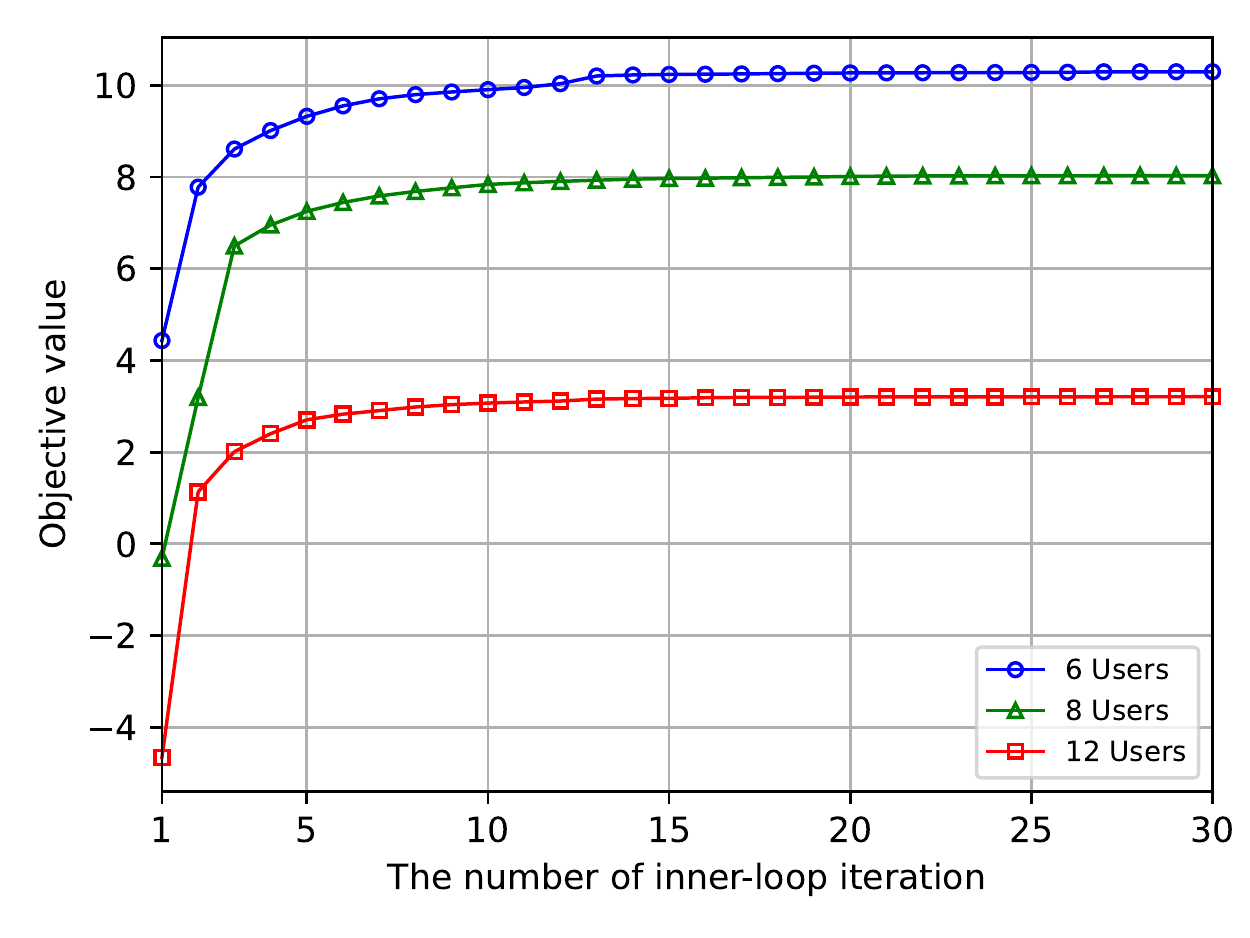}}
\caption{Convergence of Algorithm~\ref{alg:overall_solution} for different numbers of users with 4 UAVs and 4 subcarriers.}
\label{fig:converge}
\end{figure}

Fig.~\ref{fig:converge} presents the convergence of the proposed PDLIO algorithm for different numbers of users with $M=4$ and $N=4$. In Fig.~\ref{fig:converge}~(a), we evaluate the convergence of the outer-loop iteration. As can be observed,  the maximum constraint violation $\max\limits_{k,m,n} \bar{c}_{k,m,n}^{L}(1-\bar{c}_{k,m,n}^{L})$ decreases with the iteration and falls below $10^{-4}$ within 30 iterations for all settings. As the number of users increases, more iterations are needed to tackle more blockage constraints and optimization variables.
The convergence of the outer-loop iterations indicates that appropriate values of the  multipliers are obtained to ensure that the obtained user-UAV-subcarrier association variables in $\bar{\mathbf{C}}^{L}$  are binary. Moreover, the convergence of the inner-loop iteration is shown in Fig.~\ref{fig:converge}~(b). It can be observed that the inner-loop converges within 20 iterations for all settings. The convergence of the outer-loop iteration and inner-loop iteration guarantees the convergence of the proposed algorithm, and ensures the
feasibility of the obtained suboptimal solution.

%
%

\begin{figure}[t]
	\begin{center}
		\includegraphics[width=0.45 \linewidth]{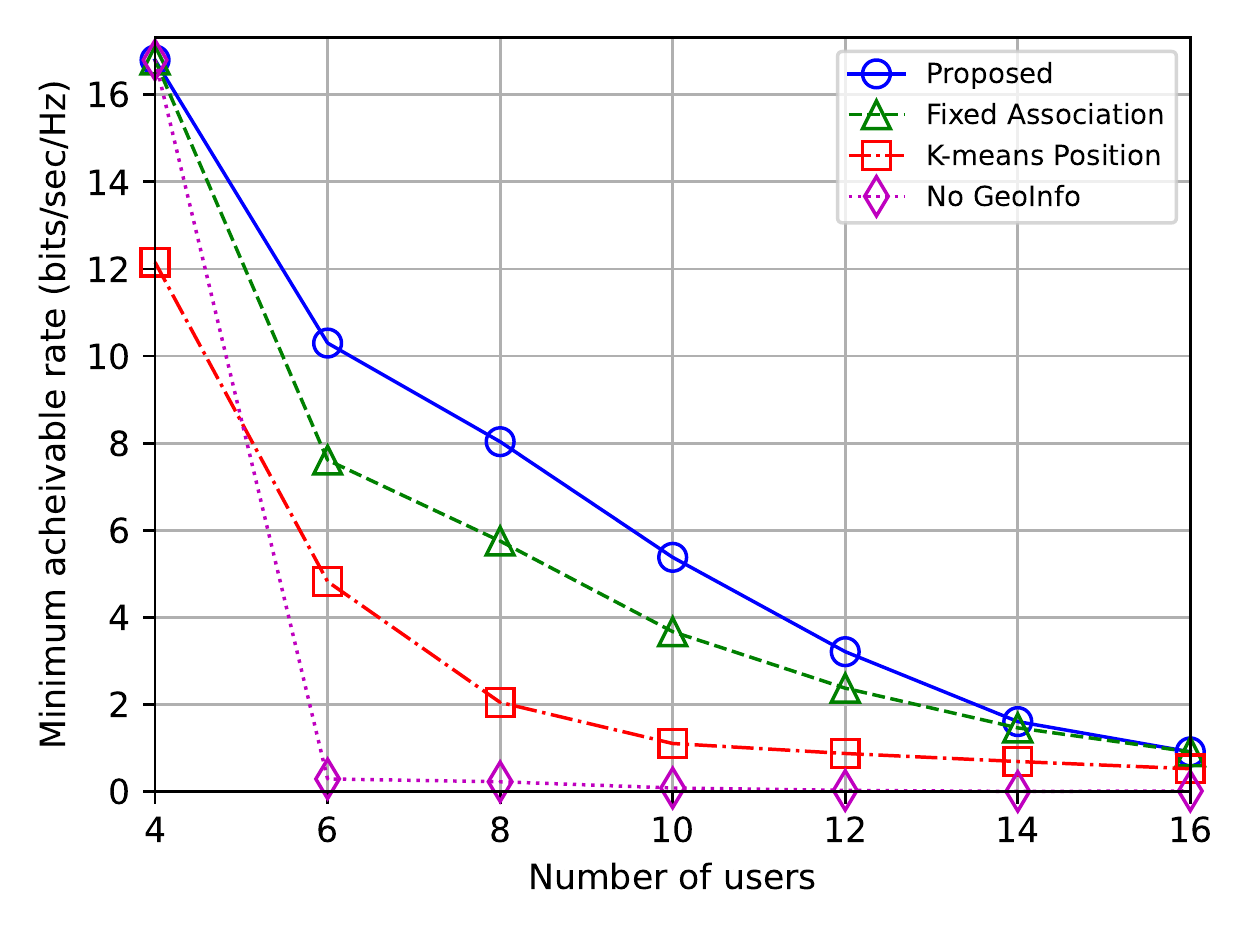}
		\caption{Minimum achievable rates for different schemes versus the number of users with 4 UAVs and 4 subcarriers.}
		\label{fig:RvsUsers}
	\end{center}
\end{figure}

Fig.~\ref{fig:RvsUsers} compares the minimum achievable rates for different schemes versus the number of users $K$ with $M=4$ and $N=4$. It can be observed that the proposed solution outperforms all other benchmark schemes. As the number of users increases, the minimum achievable rate decreases. The reason is as follows. First, 
as $K$ increases, the transmit power that can be potentially allocated to each user is reduced, leading to lower power of the received signals. Second, a limited number of subcarriers may lead to more severe mutual interference with more users. Last, as $K$ increases, more blocked regions are involved, and the UAVs tend to be deployed at higher altitudes to avoid signal blockage, leading to higher path loss. Besides, the proposed solution has a performance similar to ``Fixed Association" and ``No GeoInfo" for $K=M=N=4$, because each user can be served by an independent UAV employing a unique orthogonal subcarrier without mutual interference, and each UAV is deployed right above its served user without any blockage. In addition, when the number of UAVs and the available subcarrier resources are limited compared to the number of users $K$, the proposed solution achieves a performance similar to ``Fixed Association", since there is little freedom for user-UAV-subcarrier association optimization. Finally, without geographic information, ``No GeoInfo" scheme gets the worst results for all settings.  Since the blockage effect is not properly considered during the UAV positioning and resource allocation, this scheme cannot guarantee practical communication performance.

\begin{figure}[t]
	\centering
	\begin{minipage}[t]{0.45\textwidth}
		\centering
\includegraphics[width= \linewidth]{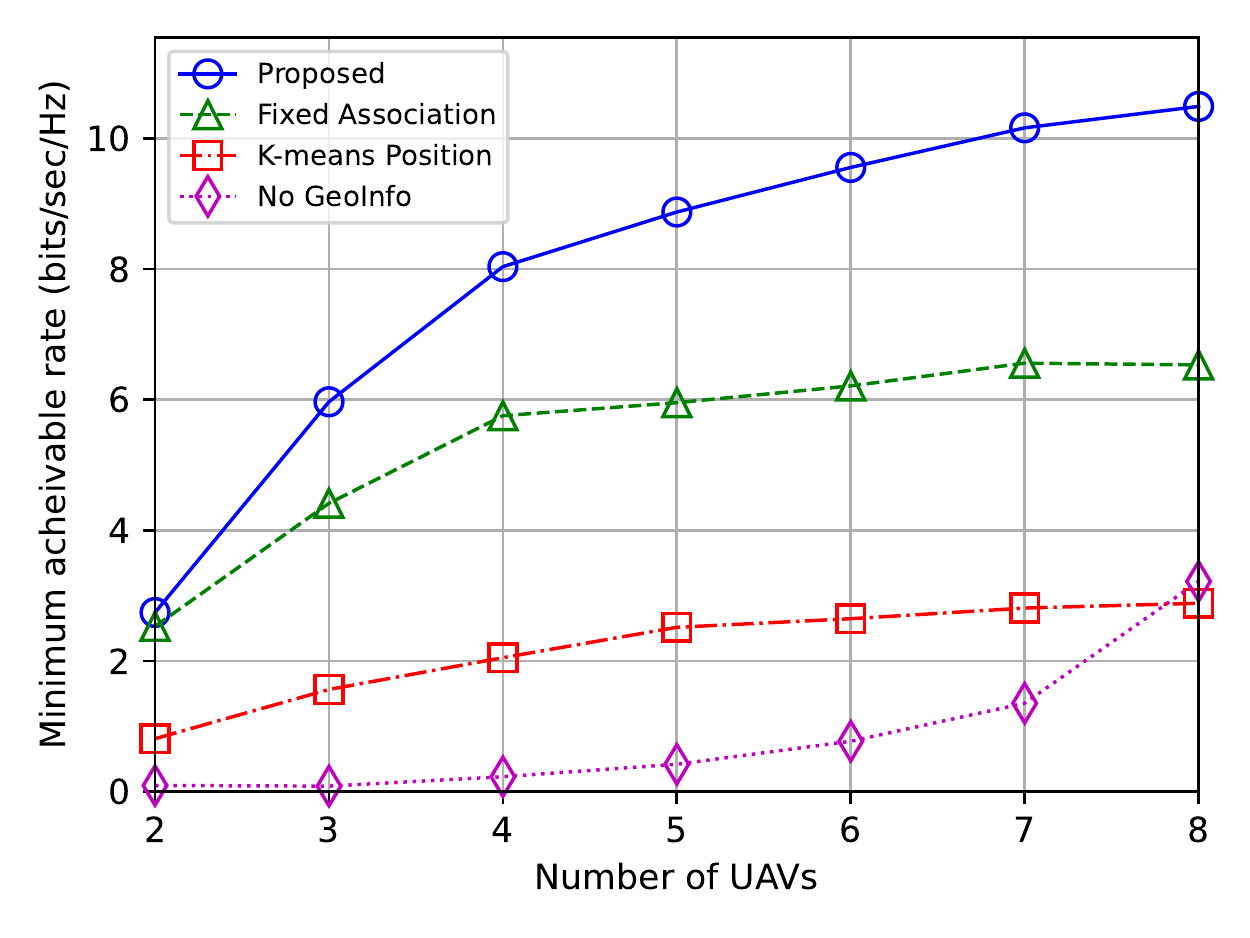}
\caption{Minimum achievable rates for different schemes versus the number of UAVs with $8$ users and $4$ subcarriers.}
\label{fig:RvsUAV}
	\end{minipage}
	\begin{minipage}[t]{0.45\textwidth}
		\centering
\includegraphics[width= \linewidth]{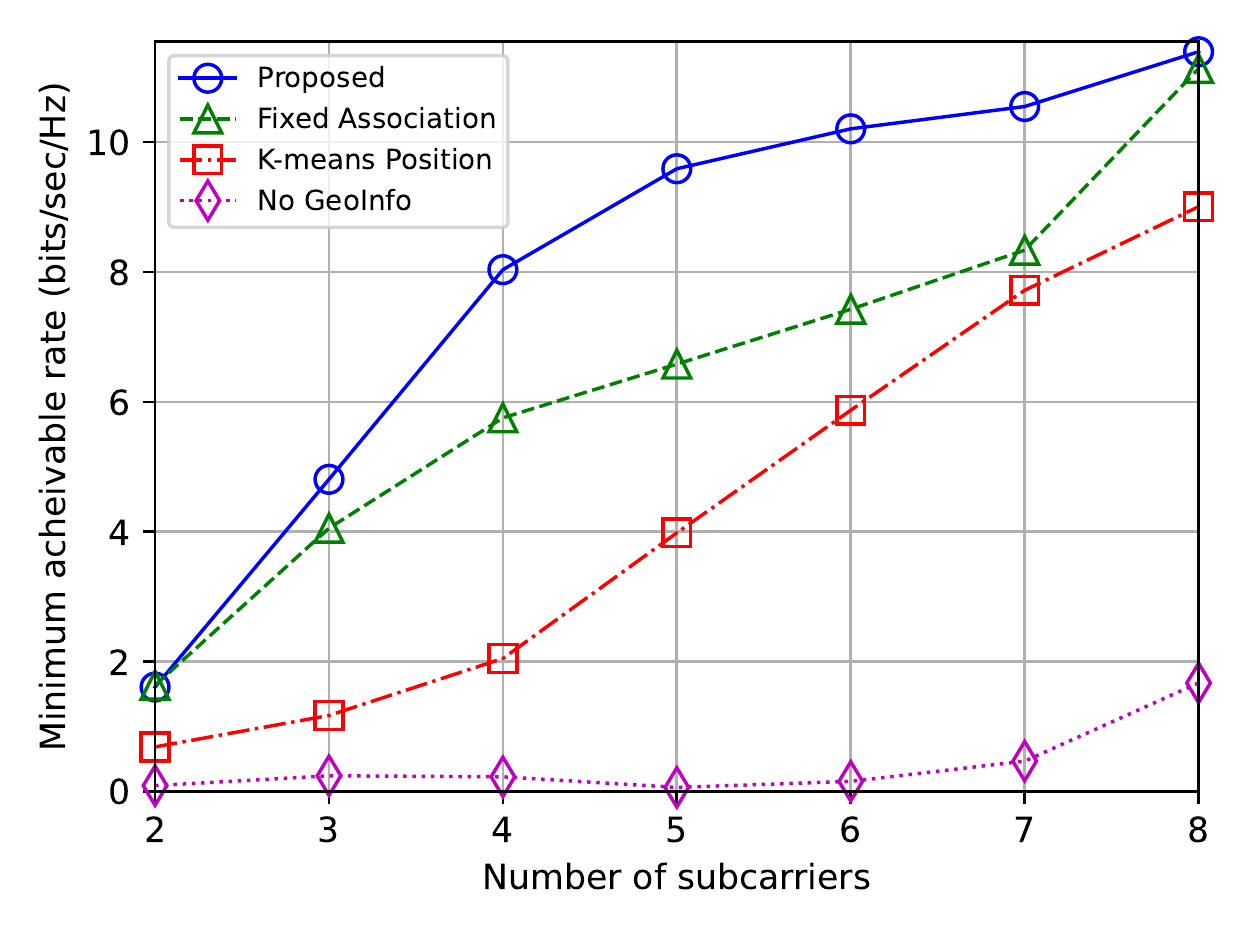}
\caption{Minimum achievable rates for different schemes versus the number of subcarriers with $8$ users and $4$ UAVs.}
\label{fig:RvsChannel}
	\end{minipage}
\end{figure}

Fig.~\ref{fig:RvsUAV} compares the minimum achievable rates for different schemes versus the number of UAVs with $K=8$ and $N=4$. As can be observed again, the proposed scheme outperforms all other benchmark schemes. As $M$ increases, the minimum achievable rate increases because there is more freedom for user association and UAV positioning. Besides, compared to ``Fixed Association'' and ``K-means Position'', the growth rate of the proposed scheme is even greater, which demonstrates the significance of  user-UAV-subcarrier association optimization and positioning optimization. Finally, although the performance of ``No GeoInfo'' scheme increases with $M$, it shows the worst practical performance, since the blockage effect is not properly addressed.


Finally, in Fig.~\ref{fig:RvsChannel} we compare the minimum achievable rates for different schemes versus the number of subcarriers  with $K=8$  and $M=4$. The proposed method still outperforms all the other benchmark schemes. Besides, we can observe that as $N$ increases, the performance gap between the proposed method and ``Fixed Association" scheme first increases and then decreases. The reason is as follows. When $N$ is small, such as $N=K/M=2$, there is little freedom for association optimization, thus leading to a small performance gap. For a large $N$, such as $N=K=8$, each user can employ a unique orthogonal subcarrier with little mutual interference. In such a case, the UAV-user-subcarrier association optimization yields little improvement in rate performance.

\section{Conclusion}\label{sec_conclusion}

In this paper, we proposed to use geographic information to characterize the LoS blockage caused by buildings, for a multi-UAV OFDMA communication system. Assisted by geographic information, a realistic channel model with blockage-aware parameters was introduced. The UAV positioning and resource allocation were then optimized to fully exploit the more favorable LoS channel conditions between UAVs and users, such that the minimum achievable rate among all the users is maximized. A penalty-based double-loop iterative algorithm was proposed to solve the challenging optimization problem. The inner-loop is to solve a penalized problem by optimizing UAV positioning sub-problem and resource allocation sub-problem in an alternating way. The outer-loop is to update multipliers to gradually decrease the violation of relaxed constraints and finally obtain a feasible solution for the original problem.
Simulation results demonstrated that the proposed scheme can guarantee practical communication performance compared to conventional LoS channel-based UAV positioning design, and achieves higher a minimum achievable rate compared to the stationary positioning and fixed association cases.



\appendices
\section{Derivation of $\nabla_{\mathbf{x}_m} g_k\left(\mathbf{x}_m \right)$,  $\nabla_{\mathbf{x}_m}\alpha_k\left(\mathbf{x}_m \right)$, and $\nabla_{\mathbf{x}_m}\beta_k\left(\mathbf{x}_m \right)$}~\label{Appendix_gradient} 

First, $\nabla_{\mathbf{x}_m} \left( \frac{\min_{q\in \mathcal{Q}} \left\{ d_{k,q}(\mathbf{x}_m) \right\}}{\|\mathbf{x}_m-\mathbf{u}_k\|} \right)$ is derived as 
\begin{equation} \notag 
\begin{aligned}
\nabla_{\mathbf{x}_m} \left( \frac{\min_{q\in \mathcal{Q}} \left\{ d_{k,q}(\mathbf{x}_m) \right\}}{\|\mathbf{x}_m-\mathbf{u}_k\|} \right)&= \frac{\nabla_{\mathbf{x}_m} \min\limits_{q\in \mathcal{Q}} \left\{ d_{k,q}(\mathbf{x}_m) \right\} }{ \|\mathbf{x}_m-\mathbf{u}_k\|   } - \frac{ \min\limits_{q\in \mathcal{Q}} \left\{ d_{k,q}(\mathbf{x}_m) \right\}  \nabla_{\mathbf{x}_m} \|\mathbf{x}_m-\mathbf{u}_k\|}{ \|\mathbf{x}_m-\mathbf{u}_k\|^2   }\\
&= \frac{\mathbf{a}_{k}(\mathbf{x}_m) }{ \|\mathbf{x}_m-\mathbf{u}_k\|   } - \frac{ \min\limits_{q\in \mathcal{Q}} \left\{ d_{k,q}(\mathbf{x}_m) \right\}  \nabla_{\mathbf{x}_m} \|\mathbf{x}_m-\mathbf{u}_k\|}{ \|\mathbf{x}_m-\mathbf{u}_k\|^2   } \\
&=\frac{\mathbf{a}_{k}(\mathbf{x}_m) }{ \|\mathbf{x}_m-\mathbf{u}_k\|   }-\frac{ (\mathbf{a}_{k}^\mathrm{T}(\mathbf{x}_m)\mathbf{x}_m-b_{k}(\mathbf{x}_m))(\mathbf{x}_m-\mathbf{u}_k) }{\|\mathbf{x}_m-\mathbf{u}_k\|^3},
\end{aligned}
\end{equation}
with 
$\left( \mathbf{a}_{k}(\mathbf{x}_m), b_{k}(\mathbf{x}_m) \right)= \arg \min\limits_{\mathbf{a}_{k,q,i},b_{k,q,i}} \left\{ d_{k,q}(\mathbf{x}_m)| q\in \mathcal{Q}\right\}$.

Furthermore, $\nabla_{\mathbf{x}_m} s(\mathbf{x}_m,\mathbf{u}_k)$ is derived as 
\begin{equation} \notag 
	\small 
\begin{aligned}
\nabla_{\mathbf{x}_m} s(\mathbf{x}_m,\mathbf{u}_k) &= \nabla_{\mathbf{x}_m} \left( \frac{1}{1+\exp \left(-\eta \frac{\min_{q\in \mathcal{Q}} \left\{ d_{k,q}(\mathbf{x}_m)\right\}}{\|\mathbf{x}_m-\mathbf{u}_k\|}  \right)} \right)= 
-\frac{ \nabla_{\mathbf{x}_m} \exp \left(-\eta \frac{\min_{q\in \mathcal{Q}} \left\{ d_{k,q}(\mathbf{x}_m)\right\}}{\|\mathbf{x}_m-\mathbf{u}_k\|}  \right)}{\left(1+\exp \left(-\eta \frac{\min_{q\in \mathcal{Q}} \left\{ d_{k,q}(\mathbf{x}_m) \right\}}{\|\mathbf{x}_m-\mathbf{u}_k\|}  \right)\right)^2} \\
&=\eta s(\mathbf{x}_m,\mathbf{u}_k)(1-s(\mathbf{x}_m,\mathbf{u}_k))  \nabla_{\mathbf{x}_m} \left( \frac{\min_{q\in \mathcal{Q}} \left\{ d_{k,q}(\mathbf{x}_m) \right\}}{\|\mathbf{x}_m-\mathbf{u}_k\|} \right).
\end{aligned}
\end{equation}

Therefore, $	\nabla_{\mathbf{x}_m}\alpha_k\left(\mathbf{x}_m \right)$ and $	\nabla_{\mathbf{x}_m}\beta_k\left(\mathbf{x}_m \right)$ are given by
\begin{equation}
\begin{aligned}
&\nabla_{\mathbf{x}_m}\alpha_k\left(\mathbf{x}_m \right)= (\alpha_1-\alpha_2) \nabla_{\mathbf{x}_m} s(\mathbf{x}_m^{{l}},\mathbf{u}_k)  = \eta(\alpha_1-\alpha_2)s(\mathbf{x}_m,\mathbf{u}_k)\left(1-s(\mathbf{x}_m,\mathbf{u}_k)\right) \times \\ &~~~~~~~~~~~~~~~~~~~~~~~~~~~~\frac{\left(\|\mathbf{x}_m-\mathbf{u}_k\|^2 \mathbf{a}_{k}(\mathbf{x}_m)  -(\mathbf{a}_{k}^\mathrm{T}(\mathbf{x}_m)\mathbf{x}_m-b_{k}(\mathbf{x}_m))(\mathbf{x}_m-\mathbf{u}_k) \right)}{\|\mathbf{x}_m-\mathbf{u}_k\|^3}, 
\end{aligned}
\end{equation}
\begin{equation}
\begin{aligned}
&\nabla_{\mathbf{x}_m}\beta_k\left(\mathbf{x}_m \right)= (\beta_1-\beta_2) \nabla_{\mathbf{x}_m} s(\mathbf{x}_m^{{l}},\mathbf{u}_k)  = \eta(\beta_1-\beta_2)s(\mathbf{x}_m,\mathbf{u}_k)\left(1-s(\mathbf{x}_m,\mathbf{u}_k)\right) \times \\ &~~~~~~~~~~~~~~~~~~~~~~~~~~~~\frac{\left(\|\mathbf{x}_m-\mathbf{u}_k\|^2 \mathbf{a}_{k}(\mathbf{x}_m)  -(\mathbf{a}_{k}^\mathrm{T}(\mathbf{x}_m)\mathbf{x}_m-b_{k}(\mathbf{x}_m))(\mathbf{x}_m-\mathbf{u}_k) \right)}{\|\mathbf{x}_m-\mathbf{u}_k\|^3}. 
\end{aligned}
\end{equation}

Finally, the derivation of $\nabla_{\mathbf{x}_m} g_k\left(\mathbf{x}_m \right)$ is given as follows:
\begin{equation}\label{appendix_gradientG}\notag 
\begin{aligned}
&\nabla_{\mathbf{x}_m} g_k\left(\mathbf{x}_m \right)=\nabla_{\mathbf{x}_m} \left( \frac{\beta_k(\mathbf{x}_m)}{\|\mathbf{x}_m-\mathbf{u}_k\|^{\alpha_k(\mathbf{x}_m)}} \right) = \frac{\nabla_{\mathbf{x}_m} \beta_k(\mathbf{x}_m)}{\|\mathbf{x}_m-\mathbf{u}_k\|^{\alpha_k(\mathbf{x}_m)}} - \frac{ \beta_k(\mathbf{x}_m) 
	\nabla_{\mathbf{x}_m}  \|\mathbf{x}_m-\mathbf{u}_k\|^{\alpha_k(\mathbf{x}_m)}}
{\|\mathbf{x}_m-\mathbf{u}_k\|^{2\alpha_k(\mathbf{x}_m)}},
\end{aligned}
\end{equation}
in which
\begin{equation}\label{appendix_gradient_1}\notag 
\begin{aligned}
&\nabla_{\mathbf{x}_m}\|\mathbf{x}_m-\mathbf{u}_k\|^{\alpha_k(\mathbf{x}_m)}  = \nabla_{\mathbf{x}_m}  \exp\left( \alpha_k(\mathbf{x}_m) \log (\|\mathbf{x}_m-\mathbf{u}_k\|) \right) \\
&= \|\mathbf{x}_m-\mathbf{u}_k\|^{\alpha_k(\mathbf{x}_m)} \left( 
\nabla_{\mathbf{x}_m} \alpha_k(\mathbf{x}_m) \cdot \log (\|\mathbf{x}_m-\mathbf{u}_k\|) +  \frac{ \alpha_k(\mathbf{x}_m)\cdot \left(\mathbf{x}_m-\mathbf{u}_k\right) }{\|\mathbf{x}_m-\mathbf{u}_k\|^2}
\right).
\end{aligned}
\end{equation}
Therefore, we have 
\begin{equation}
\begin{aligned}
\nabla_{\mathbf{x}_m} g_k\left(\mathbf{x}_m \right)=&
- 
g_k\left(\mathbf{x}_m \right) \cdot        \nabla_{\mathbf{x}_m}\alpha_k\left(\mathbf{x}_m \right) \cdot \log  (\|\mathbf{x}_m-\mathbf{u}_k\|) 
\\&-g_k\left(\mathbf{x}_m \right) \cdot\frac{\alpha_k\left(\mathbf{x}_m \right) \cdot (\mathbf{x}_m-\mathbf{u}_k) }{\|\mathbf{x}_m-\mathbf{u}_k\|^2}   +  \frac{\nabla_{\mathbf{x}_m}\beta_k\left(\mathbf{x}_m \right) }{\|\mathbf{x}_m-\mathbf{u}_k\|^{\alpha_k\left(\mathbf{x}_m \right)} }. 
\end{aligned}
\end{equation}

\bibliographystyle{IEEEtran} 
\bibliography{IEEEabrv,yi_multiUAV}
\end{document}